\documentclass[a4paper, fleqn, usenatbib]{mnras}
\usepackage{flushend}
\usepackage{newtxtext,newtxmath}
\usepackage{soul}
\usepackage{hyperref}
\usepackage{float}
\usepackage{ulem} 
\usepackage{graphicx}
\usepackage{subfig, caption}
\usepackage[T1]{fontenc}
\usepackage{ae,aecompl}
\usepackage{amsmath}
\usepackage{amssymb}
\usepackage{scalerel}
\usepackage{tikz}
\usetikzlibrary{svg.path}
\definecolor{orcidlogocol}{HTML}{A6CE39}
\tikzset{
	orcidlogo/.pic={
		\fill[orcidlogocol] svg{M256,128c0,70.7-57.3,128-128,128C57.3,256,0,198.7,0,128C0,57.3,57.3,0,128,0C198.7,0,256,57.3,256,128z};
		\fill[white] svg{M86.3,186.2H70.9V79.1h15.4v48.4V186.2z}
		svg{M108.9,79.1h41.6c39.6,0,57,28.3,57,53.6c0,27.5-21.5,53.6-56.8,53.6h-41.8V79.1z M124.3,172.4h24.5c34.9,0,42.9-26.5,42.9-39.7c0-21.5-13.7-39.7-43.7-39.7h-23.7V172.4z}
		svg{M88.7,56.8c0,5.5-4.5,10.1-10.1,10.1c-5.6,0-10.1-4.6-10.1-10.1c0-5.6,4.5-10.1,10.1-10.1C84.2,46.7,88.7,51.3,88.7,56.8z};
	}
}
\newcommand\orcidicon[1]{\href{https://orcid.org/#1}{\mbox{\scalerel*{
				\begin{tikzpicture}[yscale=-1,transform shape]
				\pic{orcidlogo};
				\end{tikzpicture}
			}{|}}}}
\title[tSZ effect in IGM]{Implications of baryon-dark matter interaction on IGM temperature and tSZ effect with magnetic field}
\author[Pandey, Malik \& Seshadri]{
	Arun Kumar Pandey\orcidicon{0000-0002-1334-043X}$^{1}$,
	\thanks{E-mail: arunp77@gmail.com,}
	Sunil Malik  \orcidicon{0000-0003-4147-626X} $^{1, a}$,
	\thanks{E-mail: sunil@physics.du.ac.in ($^a$ corresponding author)}
	T. R. Seshadri \orcidicon{0000-0002-0941-4942}$^1$
	\thanks{E-mail: trs@physics.du.ac.in}
	\\
	$^{1}$ Department of Physics and Astrophysics, University of Delhi, New Delhi-110 007, India
}
\date{\today}
\begin{document}
	\label{firstpage}
	\pagerange{\pageref{firstpage}--\pageref{lastpage}}
	\maketitle
\begin{abstract}
We show that the combined effect of cosmic magnetic field and a possible non-standard interaction between baryons and dark matter has interesting consequences on the thermal Sunyaev$-$Zel${'}$dovich (tSZ) effect depending on the temperature and the ionization state of the intergalactic medium. The drag force between the baryons and dark matter due to the relative velocity between them, and their temperature difference results in heat transfer between these two species.  At the same time the ambipolar diffusion and the decaying magnetic turbulence tends to heat up the baryons. This interplay of these two processes give rise to different evolution histories of the thermal and ionization state of the universe and hence influences the CMB spectrum at small scales through the tSZ effect. In this work, we have computed the evolution of the temperature, ionization fraction and the y-parameter of the CMB for  different strengths of the magnetic field and the interaction cross-section. We note that an interaction cross-section of $\sigma_0=10^{-42}$ m$^{2}$ (with a magnetic field strength of $3.0$ nG) reduces the y-parameter by an order of magnitude as compared to the case with similar strength of magnetic field but where there is no such interaction between the baryons and dark matter.
\end{abstract}
\section{Introduction}\label{sec:intro}
Observations indicate the existence of magnetic fields of varying strengths on  different scales in the universe (\cite{Kronberg:1977pp, Kronberg:1982pj, Welter:1984jj,	Mestel:1984rb, Rees:1987QJ, Watson:1991am, You:2003Ac, Beck:2005ra, Kronberg:2008ml, Bernet:2008ml, Bernet:2010ml, Bernet:2012fm, Hammond:2012ar, Bhat:2013kp, vacca:2018vg, Malik:2017tr}). While several theoretical models have been proposed in the literature to explain the origin of the magnetic fields at cosmological length scales, none of these mechanisms can be considered fully satisfactory in all respects. There are primarily two generation mechanisms, discussed in the literature to explain the origin of large scale magnetic fields. The first one operates during the period of large-scale structure formation and other one operates in the early universe, typically during inflation or during phase transitions. The seed fields produced in the latter one are subsequently amplified by astrophysical processes like dynamo mechanism and flux freezing in collapsed objects. In fact, it would be difficult to explain the presence of magnetic fields in voids without resorting to mechanisms that ascribe the origin to processes like inflation in the early universe. This is due to the small coherence length scales of magnetic fields that are produced purely by astrophysical processes (\cite{Furlanetto:2001gx, Bertone:2006vb}). Thus the cosmologically generated magnetic fields from processes in the early universe become almost imperative to explain the presence of these fields at a sufficiently large length scale (\cite{Hogan:1983cj, turner:1988mw, Quashnock:1989sl,  Dolgov:1993sj, Gasperini:1995vg, Joyce:1997uy,	Widrow:2002lm, Tashiro:2012mf,  vachaspati:2012tv,   Naoz2013, Pandey:2015kaa, Anand:2017zpg, Sharma:2018kgs, Jagannathan:2019kyt}) (for details see review and reference therein (\cite{Grasso:2000wj, Giovannini:2003yn, Subramanian:2015lua})). The magnetic fields generated during the early universe are referred to as primordial magnetic fields (PMFs). These PMFs are believed to be one of the possible precursors of large scale magnetic fields in the intergalactic medium (IGM) (\cite{Joyce:1997uy, vachaspati:2012tv}). 

The strength and properties of the magnetic field depends on the generation mechanism and are characterized by the present-day strength $B_\lambda$ at a length scale $\lambda$. The upper bound on $B_\lambda$ from the cosmic microwave background (CMB) observations of PLANCK are of the order of nano-gauss (nG) (\cite{Trivedi:2012ssp, Trivedi:2013wqa, Ade:2015cva}). These bounds were calculated after considering various effects of magnetic fields on CMB temperature and polarization, the ionization history, magnetically induced non-Gaussianities and magnetically induced violation of statistical isotropy. However, from the observations of the $\gamma-$rays emitted from distant blazars,  indicate that the lower bound on the strength of magnetic fields is $\sim 10^{-16}$ G in voids (\cite{Ando:2010rb, Neronov:1900zz, Essey:2010nd, Takahashi:2013lba,  Chen:2014rsa, Anand:2017zpg}). The upper bound on the strength of the magnetic field from the large scale observations is inferred to be about $1.5-4.5$ nG using observations of the Lyman-$\alpha$ forest, halo abundance and the thermal Sunyaev$-$Zel${'}$dovich effect (tSZ) effect (\cite{Kahniashvili:2012dy}). Thomson optical depth, (\cite{Kunze:2014eka}), 21 cm anisotropies (\cite{Sethi:2004pe, Tashiro:2005ua, Shiraishi:2014mt}), cosmic shear and the galaxy surveys are also some of the important probes used to constraint the strength of the magnetic fields.

The existence of magnetic fields 
could have significantly affected the CMB, large-scale structure formation as well as the 21 cm line signal (\cite{Tashiro:2005ua, Chongchitnan:2013vpa, Bhatt:2019lwt}). These fields can induce velocity  fluctuations between the ions and the residual electrons by the Lorentz force and heat up the gas in the IGM due to frictional force between the charged and neutral particles (known as ambipolar diffusion) (\cite{Cowling:1956gt, Wasserman:1978iw, Kim:1994zh, Munoz:2015bk}). Further, if PMFs affected  structure formation, it may be expected that  their imprints  may be left on the temperature and polarization anisotropies, and the thermal spectrum of the CMBR (\cite{Sethi:2004pe}).  It has been shown in references (\cite{Tashiro:2011hn, Shaw:2012la}) that in the presence of the magnetic fields, an early formation of dark halos could be possible in the galaxy clusters which cause Sunyaev$-$Zel$^{'}$dovich (SZ) effect in CMB and amplifies the angular power spectrum of the CMB temperature anisotropy on small scale. However, authors of the reference (\cite{Minoda:2017iob}) have investigated the thermal Sunyaev$-$Zel$^{'}$dovich (tSZ) effect in the IGM due to PMFs. They showed that in the presence of a random Gaussian PMFs, heating efficiency of the ambipolar diffusion is not spatially homogeneous and hence a fluctuations of the IGM gas temperature arise. In addition to the above mentioned fluctuations, matter density fluctuations are also generated in presence of the magnetic fields. This results in an observable anisotropic Compton $y-$parameter on the sky plane resulting in CMB temperature anisotropy that may be generated due to the tSZ effect. Apart from this magnetic heating of the IGM, if baryon-DM (BDM) interaction is considered, temperature of baryons may be affected because of a relative velocity and difference in the temperature of the two fluids. In the present work, we consider the possible role of a magnetic fields in determining the thermal and ionization history of the universe in the post-recombination epoch in presence of such a BDM interaction. The BDM interaction could result in the drag between the DM and baryons after the recombination era till redshift $z \sim 10$, when structure formation has just started. Further, we have also propagated these effects in the tSZ effect in IGM.

This article is organized as follows: In section (\ref{sec:1}), we have briefly summarized the role of the magnetic fields on the evolution of density fluctuations in IGM. In this section we have also explained the thermal evolution of the baryon in the standard scenario and in the presence of magnetic fields along with its decay via ambipolar diffusion and turbulent decay. Section (\ref{sec:bar-DM-main}), contains the details of the baryon-dark matter (BDM) interactions and it's effect on the thermal history of the universe after the recombination era. We have discussed the  results in section (\ref{sec-result}), where we have also computed the y-parameter of the tSZ effect (subsection (\ref{sec:sub-sz})). Finally, we have concluded our significant finding in section (\ref{sec-conl}). In the present work, we have considered a standard flat concordance model of cosmology with the following parameter $h=0.678$ ($H_0=h\times 100$ km-s$^{-1}$ Mpc), $h^2\Omega_b=0.022$, $h^2\Omega_m=0.142$ and $h^2\Omega_\Lambda=0.318$ (\cite{PDG:2012ba, Ade:2015xua}), Boltzmann constant $k_B=1.381\times 10^{-23}$ J-K$^{-1}$ and speed of light $c=2.998\times10^8$  m-s$^{-1}$.
\section{standard thermal history of baryon and the magnetic fields}
\label{sec:1}
Prior to recombination, the matter in the universe was perfectly conducting. Hence, in this era, we can consider the evolution of magnetic fields as $B\propto 1/a^2$ on scales large than magnetic Jeans scale (defined by the relation $\lambda_J^{-1}=k_J=\sqrt{8\pi \rho_m G}/\text{v}_A$, where $\text{v}_A=B_0/\sqrt{4\pi \rho_b}$ is the Alf\'ven velocity) (\cite{Sethi:2004pe}). Since, the comoving value of magnetic Jeans length $\lambda_J^c=\lambda_J /a$, remains unchanged with time, any scale which is linear/nonlinear remains linear/nonlinear even after the recombination epoch (provided vacuum energy start dominating). After the recombination epoch, the universe becomes almost neutral, and the number density of electrons sharply drops and CMB photons decouple. Therefore, the universe is no longer perfectly conducting and at a length scale $\lambda_J^c<l$, magnetic fields and the density perturbation grow linearly (\cite{Wasserman:1978iw, Gopal:2003kr}). It has been shown that, in this regime, magnetic fields decay in the IGM via ambipolar diffusion (\cite{Cowling:1956gt, Shu:1992fh}). The length scale, below which magnetic fields damp due to the radiative viscosity in the pre-recombination era, is known as Alf\'ven wave damping length scale and it is defined by the relation $l_{\rm D}=2\pi k_D^{-1}$, where, (\cite{Jedamzik:1998kk})
\begin{equation} \label{eq:mag-kd}
\frac{k_D}{2 \pi \,{\rm Mpc}^{-1}} = \left[1.32\times 10^{-3} \left(\frac{B_0(t)}{1 {\rm nG}}\right)^2 \left(\frac{\Omega_b h^2}{0.02}\right)^{-1}\left(\frac{\Omega_m h^2}{0.15} \right)\right]^{\frac{-1}{(n_{\small B}+5)}}\, .
\end{equation}
Here $n_{\small B}$ is the spectral index of the magnetic fields and, 
$B_0(t) =B_0(1+z)^2$ is the comoving magnetic field. It has been pointed out in references (\cite{Kunze:2014eka, Ade:2015cva}), that the effect of scalar and vector modes for a field of nG strength with positive spectral index $n_B>0$, can dominate over the primary CMB anisotropy at small angular scale. However, the tensor modes for fields with few nG and a nearly scale invariant spectrum (for which $n_B=-2.9$), can dominate over large length scales. In the present work, we have only considered magnetic spectrum with $n_B<0$ and ignored contributions from the scalar and vector modes and only have focused on the dissipation of the magnetic fields in the post recombination era. After this era, the radiative viscosity decreases rapidly and therefore, at scales smaller than the magnetic Jeans scale, non-linear effects lead to a decaying magnetohydrodynamic (MHD) turbulence (\cite{Jedamzik:1998kk, Subramanian:1997gi}). Therefore, at smaller scales, energy density associated with magnetic fields decay because of two effects, (i) ambipolar diffusion and (ii) turbulent decay of magnetic fields after recombination epoch. These processes will then affect the thermal and ionization history of the universe.

In this section, we have first discussed the effect of magnetic fields on the baryon density fluctuations, $\delta_{b}$. We have introduced the evolution of baryon temperature and ionization fraction in the standard picture. Then we have given a brief overview of the decay of magnetic fields in the early universe  via ambipolar and turbulent decay.

\subsection{Evolution of temperature and density fluctuations of baryons}
 A number of factors influence the thermal evolution of baryons. In addition to the adiabatic cooling due to cosmic expansion 
 and the adiabatic expansion or compression of density perturbation in baryons (${\dot T_b}=-T_bd(\ln(1+\delta_b))/dt$, where $T_b$ is the baryon temperature), the processes that can influence their thermal evolution are Compton scatterring from CMB photons, free-free cooling (bremsstrahlung), the collisional excitation cooling, the recombination cooling and the collisional ionization cooling. The inclusion of processes involving magnetic field introduces further pathways of temperature evolution of baryons via ambipolar diffusion and decay of magnetic turbulence. The rate of change of baryon temperature is proportional to the difference in temperature of baryons and the CMB photons. The proportionality factor is denoted by $\Gamma_C$, and is defined as
\begin{eqnarray}
\Gamma_{C}= \frac{8 \sigma_T\, x_e \rho_{\rm CMB}}{3(1 + x_e + f_{\rm He}) m_e}, \nonumber
\end{eqnarray}
where $x_e=n_e/n_H$ is the electron fraction, $f_{\rm He}$ is the helium fraction, $\rho_{\rm CMB}=a_r T_{\rm CMB}^4$ ($a_r=7.565\times 10^{-16}$ J-m$^{-3}$-K$^{-4}$ is the radiation constant) and $\sigma_T=6.652\times 10^{-29}$ m$^2$, is the Thomson scattering cross section. The temperature evolution due to this process is expressed as ${\dot T_b}=-\Gamma_c (T_b-T_{CMB})$. The helium ionization fraction is given by,  $f_{\rm He}= \frac{y_p}{N_{\rm tot}(1-y_p)}$ with $N_{\rm tot}=3.971$ and  $y_p=0.21$. The baryon cooling by different processes other than BDM, is characterized by the rate of change of thermal energy density in the baryons and is collectively denoted by $\Gamma_{\rm cool}$ and it is defined as
\begin{eqnarray}
\Gamma_{\rm cool}=\frac{x_e n_H }{1.5 k_B}\left[\Theta x_e +\Psi (1-x_e)+\eta x_e+\zeta (1-x_e)\right] \label{eq:cooling-terms}.
\end{eqnarray}
Here $\Theta$, $\Psi$, $\eta$ and $\zeta$ governs free-free cooling (bremsstrahlung), the collisional excitation cooling, the recombination cooling and the collisional ionization cooling respectively, and they are given as (\cite{Fukugita:1994mm})
\begin{eqnarray}
\Theta& = & 5.84 \times 10^{-37}\, T_{-5}^{0.5}\, [\text{J}\,\text{m}^3\,\text{s}^{-1}], \nonumber \\
\Psi & = & 7.50\times 10^{-31}  \left(1+T_{-5}\right)^{-1} \exp\left(-\frac{1.18}{T_{-5}}\right)
[\text{J}\, \text{m}^3\,\text{s}^{-1}], \nonumber \\
\eta & = & 2.06 \times 10^{-36}T_{-5}^{0.5}\left(T_{-5}\right)^{-0.2} \left(1+\left(T_{-5}\right)^{0.7}\right)^{-1}
[\text{J}\, \text{m}^3\,\text{s}^{-1}],\nonumber \\
\zeta & = & 4.02 \times 10^{-32} T_{-5}^{0.5}\left(1+T_{-5}^{-1}\right)\exp\left(-\frac{1.58}{T_{-5}}\right)\, \, 
[\text{J}\, \text{m}^3\,\text{s}^{-1}],
\end{eqnarray}
where $T_{-5}=T_b/10^5$. This process leads to a change in the baryon temperature at a rate given by ${\dot T_b}=-\Gamma_{\rm cool}(\frac{3}{2}n_Bk_B)^{-1}$. 

As pointed out earlier, the presence of magnetic fields affects the baryon temperature via ambipolar diffussion and turbulent decay. Hence we need to consider the evolution of the magnetic energy density ($B^2/8\pi$). The evolution equation for the magnetic energy density is given by:
\begin{eqnarray}
\frac{d}{dt}\left(\frac{|{\bf B}|^2}{8\pi}\right)= -4 H(t)\left(\frac{|{\bf B}|^2}{8\pi}\right)-\Gamma_{\rm heat}\, , \label{eq:mag_heat-1}
\end{eqnarray}
Note that $\Gamma_{\text {heat}}$ refers to the heating of baryons and hence draining of energy from magnetic field and
is defined as;
\begin{eqnarray}\label{eq:gam-heat}
\Gamma_{{\rm heat}} = \Gamma_{{\rm ambi}} + \Gamma_{{\rm tur}}\, .
\end{eqnarray} 
 In (\ref{eq:gam-heat}), the first term on the right hand side represents the heat dissipation per unit volume via ambipolar diffusion and second term is the decay of magnetic energy via MHD turbulent decay.
The ambipolar decay rate is defined as \cite{Sethi:2003vp}:
\begin{eqnarray}
\Gamma_{{\rm ambi}} = \frac{\rho_n}{16\pi^2\gamma \rho_b^2 \rho_i} |(\nabla\times {\bf B})\times {\bf B}|^2\,, \label{eq:ambi1}
\end{eqnarray}
where $\rho_n$, $\rho_i$ and $\rho_b$ are the mass densities of neutral, the ionized atoms and total baryon density, respectively. Also, $\gamma$ in the present scenario is given by (\cite{Shang:2001df, Schleicher:2008aa})
\begin{eqnarray}
\gamma = \frac{\frac{1}{2}n_{\rm H} \langle \sigma v \rangle_{{\rm H}^+, {\rm  H}}+\frac{4}{5}n_{\rm He}\,\langle \sigma v \rangle_{{\rm H}^+, {\rm He}}}{m_{\rm H} \, (n_{\rm H}+4 n_{\rm He})}\, , 
\end{eqnarray}
where $n_{\rm H}$ and $n_{\rm He}$ are the number density of hydrogen and the helium, respectively, and $m_{\rm H}$ is the mass of the hydrogen atom. In the present work, the effect of collisions with the electrons is neglected as its contribution in above relation is suppressed by a factor $m_e/m_{\rm H}$. The two quantities $\langle \sigma v \rangle_{{\rm H}^+, {\rm  H}}$ and $\langle \sigma v \rangle_{{\rm H}^+, {\rm He}}$ are the interaction cross-sections of collision between ${\rm H}^+$ $-$ H and ${\rm H}^+$ $-$ ${\rm He}$, respectively, which are given by (\cite{Schleicher:2008aa, Pinto:2008zn, Pinto:2008ab})
\begin{eqnarray}
\langle \sigma v \rangle_{{\rm H}^+, {\rm  H}} & = & 0.649\, T_b^{0.375}\times 10^{-15}\, {\rm m}^3 \, s^{-1}, \nonumber \\
\langle \sigma v \rangle_{{\rm H}^+, {\rm He}} & = &  [1.424+7.438\times 10^{-6} T_b-6.734\times 10^{-9} T_b^2]\nonumber \\
&\times &  10^{-15}\, {\rm m}^3 \, s^{-1} .
\end{eqnarray}
The decay of magnetic energy by MHD turbulence is formulated via numerical simulations and we have adopted the description of magnetohydrodynamic decay of MHD modes as given in \cite{Sethi:2004pe}. The rate of magnetic energy via turbulent decay is given by
\begin{eqnarray} 
\Gamma_{{\rm tur}} =  \frac{B_0^2(t)}{8\pi}\frac{3m}{2}\frac{\left[\ln\left(1+t_{\rm eddy}/t_i\right)\right]^m\, H(t)}{\left[\ln\left(1+t_{\rm eddy}/t_i\right)+\ln(t/t_i)\right]^{m+1}}\, , \label{eq:decay-1}
\end{eqnarray}
where $t$ is the cosmic time at any redshift $z$, $m$ is a parameter that depends on the magnetic spectral index, $t_{\rm decay}=k_{D}^{-1}/{\rm v}_{\rm A}$ is the dynamical timescale and $t_i=1.248\times10^{13}$ s is the time scale over which velocity perturbations are no longer damped by the large radiative viscosity after the recombination epoch. Alf\'ven velocity is defined as ${\rm v}_{\rm A}=1.5\times 10^{-5}\,\frac{c \,B_0}{10^{-9}(1+z)^2 {\rm in~ G}} (1+z)^{1/2}$. Taking all the above factors into account the evolution of baryon temperature with time can be expressed as
\begin{equation}
\frac{dT_b}{dt}=-2H\,T_b + \frac{T_b}{1+\delta_b}\frac{d \delta_b}{dt}- {\Gamma_{C}} (T_b-T_{\rm CMB})
+\frac{(\Gamma_{\rm heat}-\Gamma_{\rm cool})}{1.5k_B\, n_{\small B}}\,,
\label{eq:baryon_temp-1}
\end{equation}
where $H(t)=\dot{a}/a$ is Hubble parameter in terms of scale factor $a(t)$ (dot above the scale factor denotes the derivative respect to cosmic time $t$). Expressed in the terms of redshift, this can be written as,
\begin{eqnarray}
\frac{dT_b}{dz} & = & \frac{2T_b}{(1+z)} + \frac{T_b}{1+\delta_b}\frac{d \delta_b}{dz}+ \frac{\Gamma_{C}}{(1+z)H} (T_b-T_{\rm CMB}) \nonumber \\
&-&  \frac{(\Gamma_{\rm heat}-\Gamma_{\rm cool})}{1.5k_B\, n_{\small B}\,(1+z)H}, \label{eq:baryon_temp-1}
\end{eqnarray}
In order to solve the above equation, we need the equations for the ionization fraction, and the evolution equation for magnetic fields energy density (given in Eq.~\ref{eq:gam-heat}). The ionization equation is given by (\cite{AliHaimoud:2010dx}),
\begin{eqnarray}
\frac{dx_e}{dz} & = &  \frac{1}{H(1+z)} \Bigg[\left(n_{\small B} x_e^2\alpha_{B}-(1-x_e)\beta_{B}e^{-E_{21}/T_{\rm CMB}} \right) D \nonumber \\
& - & \gamma_e\, n_{\small B} (1-x_e)\, x_e \Bigg]
\label{electron}
\end{eqnarray}
where $\alpha_{B}$, $\beta_{B}$ are case-B recombination coefficient and photoionization rate, respectively and $\gamma_e$ is collisional coefficient. These parameters are given by
\begin{eqnarray}
\alpha_B & = & F\times 10^{-19}\frac{a T_{-4}^b}{1+cT_{-4}^d} \,\, [{\rm m}^3 \, {\rm s}^{-1}], \\
\beta_B & = & \alpha_B \left(\frac{2\pi m_e k_B T_{\rm CMB}}{h_{\rm pl}^2}\right)^{3/2}\exp\left(\frac{E_{21}}{k_B T_{\rm CMB}}\right)\, [ {\rm s}^{-1}]\, ,\\
\gamma_e & = & =0.291\times10^{-7} \times U^{0.39} \, \frac{\exp(-U)}{0.232+U}\,\,[{\rm cm}^3 {\rm s}^{-1}] \, .
\end{eqnarray}
Here $E_{21}=E_{1s}-E_{2s}=-10.2$ eV is energy of Ly$\alpha$ wavelength photon, $E_{1s}=-13.6$ eV, $U=|E_{1s}/ k_B T_b|$ and $E_{2s}=-3.4$ eV. Other parameters used are $T_{-4}=T_b/10^{4}$ K, Boltzmann constant $k_B= 1.38\times 10^{-23}{\rm m}^2\, {\rm kg}\, {\rm s}^{-2}\, {K}^{-1}$, $a=4.309$, $b=-0.616$, $c=0.670$ and $d=0.530$. The suppression factor $D$, due to the Ly-$\alpha$ photons is given by,
\begin{eqnarray}\label{eq:Dfactor}
D=\frac{1+K_H\Lambda_{2s,1s} n_{\small B} (1-x_e)}{1+K_H \, n_{\small B} (1-x_e)+K_H n_B \beta_B (1-x_e)}
\end{eqnarray}
where $K_H= \lambda_{H_{2p}^3}/8\pi H$, $\lambda_{H_{2p}}=121.568$ nm and  $\Lambda_{2s,1s}= 8.22~ {\rm s}^{-1}$ is the hydrogen two photon decay rate.
The evolution of density perturbation in baryon fluid  is accelerated since they fall into the potential wells made by dark matter perturbations. Further, magnetic field also induces perturbations in the baryon fluid via Lorentz force due to the velocity fields
(\cite{Wasserman:1978iw, Kim:1994zh, Subramanian:1997gi, Gopal:2003kr, Sethi:2004pe, Tashiro:2006uv}). In the standard picture, there is no interaction between baryons and Dark Matter particles, other than gravity. The equations governing the evolution of the density perturbations in dark matter and baryons in the presence of magnetic fields is thus given by:
\begin{eqnarray}
&&\frac{\partial^2 \delta_d}{\partial t^2} + 2 H(t) \frac{\partial \delta_d}{\partial t} - 4\pi G (\rho_d \delta_d + \rho_b \delta_b) = 0,	\label{eq:cdm-1} \\
&&\frac{\partial^2 \delta_b}{\partial t^2} + 2 H(t)\frac{\partial \delta_b}{\partial t}- 4\pi G (\rho_d \delta_d + \rho_b \delta_b) = S(t).	\label{eq:baryon-1}
\end{eqnarray}
Here, $\rho_b$ and $\rho_d$ are the baryon and cold dark matter density, and $\delta_{\rm b}$ and $\delta_{\rm d}$ are the density contrast of the baryon and cold dark matter, respectively. In Eq.~\eqref{eq:baryon-1}, $S(t)$ represents the spatially averaged contribution from the Lorentz forces due to magnetic fields, which is given as \cite{Minoda:2017iob}
\begin{eqnarray}
S(t, {\bf x}) = \frac{\nabla \cdot \left[(\nabla \times \mathbf{B}(t,\mathbf{x})) \times \mathbf{B}(t,\mathbf{x})\right]}{4 \pi \rho_\mathrm{b} (t) a^2(t)}, \label{eq:source_b-1}
\end{eqnarray}
where $B(t, {\bf x})$ is the comoving magnetic fields measured in comoving coordinates ($t, {\bf x}$). All other variables are also in comoving coordinate system. We assume that the conductivity of the fluid is infinite, and hence we can safely ignore any back reaction from the local matter distribution on the PMFs. Under these assumptions, magnetic fields evolve adiabatically and can be expressed as: ${\bf B}({t, {\bf x}})= {\bf B}_0({\bf x})/a^2(t)$ (where $B_0=|{\bf B}_0|$ is the comoving value of the PMFs). In the present work, we have evolved temperature of the baryons, DM and the  ionization fraction from redshift $z=1100$ to redshift $z=10$. At redshift $z = 1100$, density perturbation $\delta_b$ is very small ($\sim 10^{-6}$ to $10^{-5}$). Even if the initial value $\delta_b$ is taken to be zero, the source term on the right hand side of equation (\ref{eq:baryon-1}) generates the perturbation. Thus one may neglect the initial value of $\delta_b$ and hence, assume it to be zero. In such a situation, we can obtain analytical solution for $\delta_{\rm b}$ by solving equation \eqref{eq:baryon-1} by Green's function method as (\cite{Minoda:2017iob}:
\begin{eqnarray}
\delta_b = \frac{2S(t)}{15H^{2}(t)}
\left[ \left\{ 3\left(\frac{a}{a_{\rm rec}} \right)  \right. \right.
+ 2\left(\frac{a}{a_{\rm rec}} \right)^{-\frac{3}{2}}
- \left. 15 \ln \left( \frac{a}{a_{\rm rec}}\right) \right\}
\frac{\Omega_b}{\Omega_m}  \nonumber \\
+ 15 \ln \left(\frac{a}{a_{\rm rec}}\right)
+ 30 \left(1 - \frac{\Omega_b}{\Omega_m}\right)
\left(\frac{a}{a_{\rm rec}}\right)^{-\frac{1}{2}}
- \left. \left(30-\frac{25 \Omega_b}{\Omega_m}\right)  \right],
\label{eq:delta_b-1}
\end{eqnarray}
where $a_{\rm rec}\propto (1+z_{\rm rec})^{-1}$ is the scale factor corresponding to the recombination epoch.
\section{Consequences of a possible additional Non-gravitation Dark Matter-baryon interaction}
\label{sec:bar-DM-main}
It is known that in the standard cosmological evolution, dark matter starts collapsing at the epoch of matter-radiation equality. However, baryons cannot as yet, form structure due to radiation pressure, since baryons remain coupled to the photons. Due to their separate evolution, there will be a non-zero relative velocity between baryons and DM. The relative velocity will hence influence the thermal history of the baryons and DM. Eventually, CMB photons and baryons decouple at redshift $z\approx 1100$. This era is known as recombination epoch and is followed by the era of dark ages. During this era, the overdense regions grow in the baryon distribution and start collapsing into the potential wells of Dark Matter, which have developed by then, to form the first stars and galaxies in the universe. Also, the baryons remain in thermal equilibrium with the CMB photons due to mutual interactions till $z\approx 200$. During this phase, baryons and CMB photons interact through Compton scattering. In this era, DM particles are in their coolest phase due to absence of any nonlinear gravitational collapse (\cite{Barkana:2018nd}). In addition to the standard gravitational interaction between DM and baryon, it is interesting to consider the possibility of other kinds of interactions between them. The most optimistic scenario for this additional interaction between the baryon-dark matter is realized by the Rutherford-like interaction i.e., a velocity-dependent interaction cross-section (\cite{Prinz:1998ba, Spergel:2000nj, Davidson:2000hf, Chen:2002xh, Dubovsky:2004ds, Sigurdson:2004kd, Melchiorri:2007sq, Jaeckel:2010ni, McDermott:2011dy, Dolgov:2013ad, Tulin:2013yh, Tulin:2013hb, Dvorkin:2013cea, Vogel:2014jv, Dvorkin:2014cb, Berlin:2018sj}). A general form of the interaction cross-section dependent on the relative velocity is given by $\sigma({\rm v})=\sigma_0 ({\rm v}/c)^{n}$, where ${\rm v}$ is the relative velocity of the DM-baryon. For a Yukawa potential (massive Boson exchange),  we have $n=-1$ (\cite{Aviles:2011ac}), $n=-2$ belongs to the case when the DM particles have an electric dipole moment (\cite{Sigurdson:2004kd} and $n=-4$  corresponds to the case of the millicharge DM particles (\cite{McDermott:2011dy, Dolgov:2013ad}). If the scattering of the baryons and DM is effective enough before recombination, it affects the linear matter power spectrum on small scales, temperature polarization and lensing anisotropies of CMB (\cite{Dvorkin:2014cb, Tashiro:2014tsa, Barkana:2018lgd}). However, if the interaction is significant in the post-recombination era, it can lead to an anomalous behavior of baryon temperature, and hence alter the 21-cm absorption signal (\cite{Tashiro:2014tsa}), and Ly$\alpha$ radiation after the first star formation (\cite{Munoz:2015bk}). It has been shown that in the presence of primordial magnetic fields, this interaction changes the dynamics of the baryon and DM significantly for a certain choice of parameters (\cite{Bhatt:2019lwt}). There are many pieces of work, where thermal and ionization history of the IGM is studied extensively in various contexts (\cite{ Tashiro:2005ua, Tashiro:2006uv, Schleicher:2008aa, Tashiro:2011hn, Tashiro:2014tsa, Shiraishi:2014mt, Minoda:2017iob, Barkana:2018nd,  Moroi:2018vci, Bhatt:2019lwt}). 
Here we have now included the BDM interaction along with the magnetic heating of the IGM. In subsection (\ref{subsec:dm-drag}), we have discussed the heat transfer between the baryons and the DM due to non-standard interaction between them and finite relative velocity during the dark ages. Further, we add a subsection related to the magnetic fields, where we have considered a power-law spectrum and derived the ambipolar decay of the magnetic fields for this spectrum (subsection \ref{mag-nature}).
\subsection{Heat exchange due to Dark Matter-Baryon interaction} \label{subsec:dm-drag}
The implications of a possible interaction between dark matter and baryons, over and above the normal gravitational interaction, has attracted much attention and has been extensively investigated in recent years. This subsection is mostly based on work done following references ((\cite{Schleicher:2008aa, Dvorkin:2013cea, Munoz:2015bk, Boddy:2018wzy}). As discussed above, due to the relative velocity at kinematic decoupling, there will be a drag force between the DM and baryons. The drag force per unit mass exerted by the baryons on DM fluid due to baryon-DM interaction is quantified by the rate of change of the bulk velocity field of the DM fluid with respect to baryons (\cite{Dvorkin:2013cea})
\begin{eqnarray}\label{eq:drag-v}
\frac{d{\bf v}}{dt}=-{\bf v}\frac{\rho_m \sigma_0 c^{-n}\,\text{v}^{n+1}}{(m_d+m_b)} F(r),
\end{eqnarray}
where $\rho_m$ is the total matter mass density, and $m_b$, $m_d$ represent the mass of the baryon and dark matter particles, respectively. Here $r \equiv \text{v}/u_{\rm th}$ and the thermal velocity dispersion $u_{\rm th}$ is given by: $u_{\rm th}^2= \left(\frac{k_b T_b}{m_b}+\frac{k_b T_d}{m_d}\right)$. The function $F(r)$ is defined as, $F(r)\equiv {\rm erf}\left(\frac{r}{\sqrt{2}}\right)-r\,\sqrt{\frac{2}{\pi}}\, {\rm exp}\left(-\frac{r^2}{2}\right)$. We may note that $F(0)=0$ and $F(r\rightarrow \infty)=1$. Throughout our present work, we assume that all baryons and DM particles are non-relativistic. The above relation has been derived in literature for different values of $n$ for the case when there is no magnetic field (\cite{Boddy:2018wzy}). The heat transfer between the two fluids consists of two terms; the first one is the heat transfer due to the difference in temperature and the second one is due to the drag between these two fluids. Due to the drag term, the final relative velocity should tend to zero. 
Therefore energy transfer to baryons is
\begin{eqnarray}\label{eq:heat_tranfer}
\left(\frac{dQ_b}{dt}\right)=\left(\frac{dQ_b}{dt}\right)_{d \rightarrow b}+\left(\frac{dQ_b}{dt}\right)_{\rm drag~ term}.
\end{eqnarray} 
Similarly we can obtain the energy transferred to dark matter is given by just exchanging $b~ \leftrightarrow d$. 
The first term in equation (\ref{eq:heat_tranfer}) is given by (\cite{Munoz:2015bk, Munoz:2017qpy}
\begin{eqnarray}\label{eq:dqdb}
\left(\frac{dQ_{b}}{dt}\right)_{d \rightarrow b}  =  \Gamma_{d\rightarrow b}(T_d-T_b)
\end{eqnarray}
where,
\begin{eqnarray}
\Gamma_{d\rightarrow b} & = & \frac{\rho_d\mu_b\, k_B\, c^{-n}\, \sigma_0 e^{-r^{2}/2}} {(m_d+m_b)^2}\left(\frac{k_B T_b}{m_b} + \frac{k_BT_d}{m_d} \right)^{\frac{n+1}{2}}\nonumber \,\\ 
& \times &\frac{2^\frac{5+n}{2} \Gamma\left(3+\frac{n}{2}\right)}{\sqrt{\pi}}
\end{eqnarray}
and $\mu_b\simeq m_H(n_H+4 n_{He})/(n_H+n_{He}+n_e)$ (see below equation (2) of reference \cite{Tashiro:2014tsa}). The second term in equation (\ref{eq:heat_tranfer}) is given by 
\begin{eqnarray}
\left(\frac{dQ}{dt}\right)_{\rm drag~ term}=\frac{\rho_d}{\rho_d+\rho_b}\frac{m_d\, m_b}{m_d+m_b}\, |{\rm v}\, \frac{d \textbf{v}}{dt}|
\end{eqnarray}
Thus in Eq.\eqref{eq:heat_tranfer}, the first term  represents the baryonic cooling due to its interaction with the DM and the second term represents the heating due to the drag term. The relative velocity between the dark matter and baryon, produces friction between the two fluid which is responsible for the drag term. At the decoupling epoch ($z\sim 1100$), the DM temperature is much lower than the baryons as the baryons are electromagnetically coupled to CMB photons. Hence at this epoch, $(dQ/dt)_{d\rightarrow b}<0$ (in equation (\ref{eq:dqdb})) and $(dQ/dt)_{b\rightarrow d}>0$ .
\subsection{Magnetic field spectrum and decay coefficients} \label{mag-nature}
Under the assumption that, magnetic field evolves adiabatically in the early universe and that it is statistically homogeneous and isotropic, the two-point correlation of the magnetic fields is defined as (\cite{Brandenburg:2018ptt})
\begin{equation}
\langle \tilde{{\bf B}}_i ({\bf k}) \, \tilde{{\bf B}}^*_j ({\bf q})\rangle = \frac{(2\pi)^3}{2}\delta_D^3({\bf k}-{\bf q})\left(\delta_{ij}-\frac{k_i k_j}{k^2}\right)\mathcal{P}_B(k)\,,
\end{equation}
where $\mathcal{P}_B(k)$ is the magnetic power spectrum and $k=|{\bf k}|$ is the comoving wave number (for details, see the appendix-(\ref{sec:mag-spec})).  Assuming a power law for the magnetic power spectrum, it can be expressed as (\cite{Sethi:2004pe}),
\begin{eqnarray}
\mathcal{P}_B(k)=
\begin{cases}
A k^{n_{\small B}},& k \leq k_D\\
0,              & k> k_D. 
\end{cases}
\end{eqnarray}
Here $k_D$ is the cut off scale (corresponding to length scale $l_D\approx 1/k_D$), below which magnetic energy undergoes Alf\'ven damping (\cite{Jedamzik:1998kk, Subramanian:1997gi, Kahniashvili:2012dy}). The amplitude $A$ can be obtained by smoothing the magnetic field and is given by (for details see the appendix (\ref{sec:mag-spec})):
\begin{eqnarray}\label{eq:Amplitude-B}
A=\frac{(2\pi)^2\, 2^{(n_B+1)/2}\, \langle B^2\rangle_\lambda}{\Gamma \left(\frac{n_B+3}{2}\,  \right)\, k_\lambda^{3+n_B}}, 
\end{eqnarray}
where, $\langle B^2\rangle_\lambda$ is the strength of magnetic fields smoothed over a length scale $\lambda$. With the above magnetic field spectrum, the ambipolar diffusion coefficient, $\Gamma_{\rm ambi}$ turns out to be
\begin{eqnarray}
\Gamma_{\rm ambi}=\frac{7}{3}\left(\frac{A}{8\pi}\right)^2 \frac{\rho_n}{16\pi^2\gamma \rho_b^2 \rho_i}  \int_{k_{\rm min}}^{k_{\rm max}} k^{2+n_{\small B}} dk  \int_{k_{\rm min}}^{k_{\rm max}} k^{4+n_B} dk.
\end{eqnarray}
The lower and upper limits are given by $k_{\rm min}=2\pi/(10 ~\text{Mpc})$ (related to largest length scale, that we have considered) and $k_{\rm max}=k_D$ (as defined in equation (\ref{eq:mag-kd})).
For a power law magnetic energy spectrum (equation (\ref{eq:decay-1})), $m$ is given by $m=2(n_{\small B}+3)/(n_{\small B}+5)$ (\cite{Olesen:1996ts, SHIROMIZU:1998ts, Jedamzik:1998kk, Christensson:2001ma, Sethi:2004pe}). 

\subsection{Resulting evolution of Baryon and DM temperature}\label{subsec-BDM-B}
Here, we would like to emphasize that, in literature, the magnetic fields and DM-baryon interactions are considered in various contexts. For example, the effect of the magnetic fields has been investigated in the context of standard (purely gravitational) interactions between the baryon-DM particles. However, in some recent works, a small additional non-gravitational BDM interaction is considered. These however, did not include magnetic field. The importance of the present work is that it demonstrates that the interplay of the combined effect of magnetic fields as well as the non-standard BDM interaction on the thermal and ionization history of IGM, results in some interesting consequences.

In addition to the terms on the RHS of Eq.~(\ref{eq:baryon_temp-1}), which drive the evolution of baryon temperature with redshift, we now also consider the contribution from the baryon-Dark Matter interaction. It also provides an additional source term for the evolution of the temperature of Dark matter as well as the evolution of the relative velocity between baryons and dark matter. The resulting evolution equations are
\begin{eqnarray}
\frac{dT_b}{dz} & = & \frac{2T_b}{(1+z)} + \frac{T_b}{1+\delta_b}\frac{d \delta_b}{dz}+ \frac{\Gamma_{C}}{(1+z)H} (T_b-T_{\rm CMB}) \nonumber \\ 
&-&  \frac{1}{1.5k_b(1+z)H}\frac{d{Q_b}}{dt}-  \frac{(\Gamma_{\rm heat}-\Gamma_{\rm cool})}{1.5k_B\, n_{\small B}\,(1+z)H}, \label{eq:baryon_temp} \\
\frac{dT_{d}}{dz}& = &\frac{2T_{d}}{(1+z)} - \frac{1}{1.5 k_B\, (1+z)H}\frac{d{Q_{d}}}{dt}\, ,
\label{eq:dm_temp} \\
\frac{d\text{v}}{dz}& =&  \frac{\text{v}}{(1+z)} + \frac{D(\text{v})}{(1+z)H} \, ,
\label{eq:vel}
\end{eqnarray}
where $T_{d}$ is the temperature of the DM. To solve these equations, we have considered following initial conditions: $x_e(z=1100)=0.057$, $T_b(z=1100)= 3000$ K, $T_d(z=1100)=0.0$ K (as dark matter temperature is negligibly small) and $v(z=1100)=10^{-4}$ m-s$^{-1}$. In the present study, we have considered BDM interaction parameter, $n=-4$ 
(although the case of $n=-2$ is also physically motivated, we have found that this does not lead to any significant effect on the thermal history of the baryons. Hence we will not be considering this case further). In the next section 
, we have discussed the results obtained by solving equations (\ref{eq:baryon_temp} - \ref{eq:vel}) using the above equations and the initial conditions described, we will study the thermal and ionization history of the interacting dark matter (DM) and baryons fluids in presence of magnetic fields during the dark ages.
\begin{figure*}
	\subfloat[]{\includegraphics[width=0.40\linewidth, keepaspectratio]{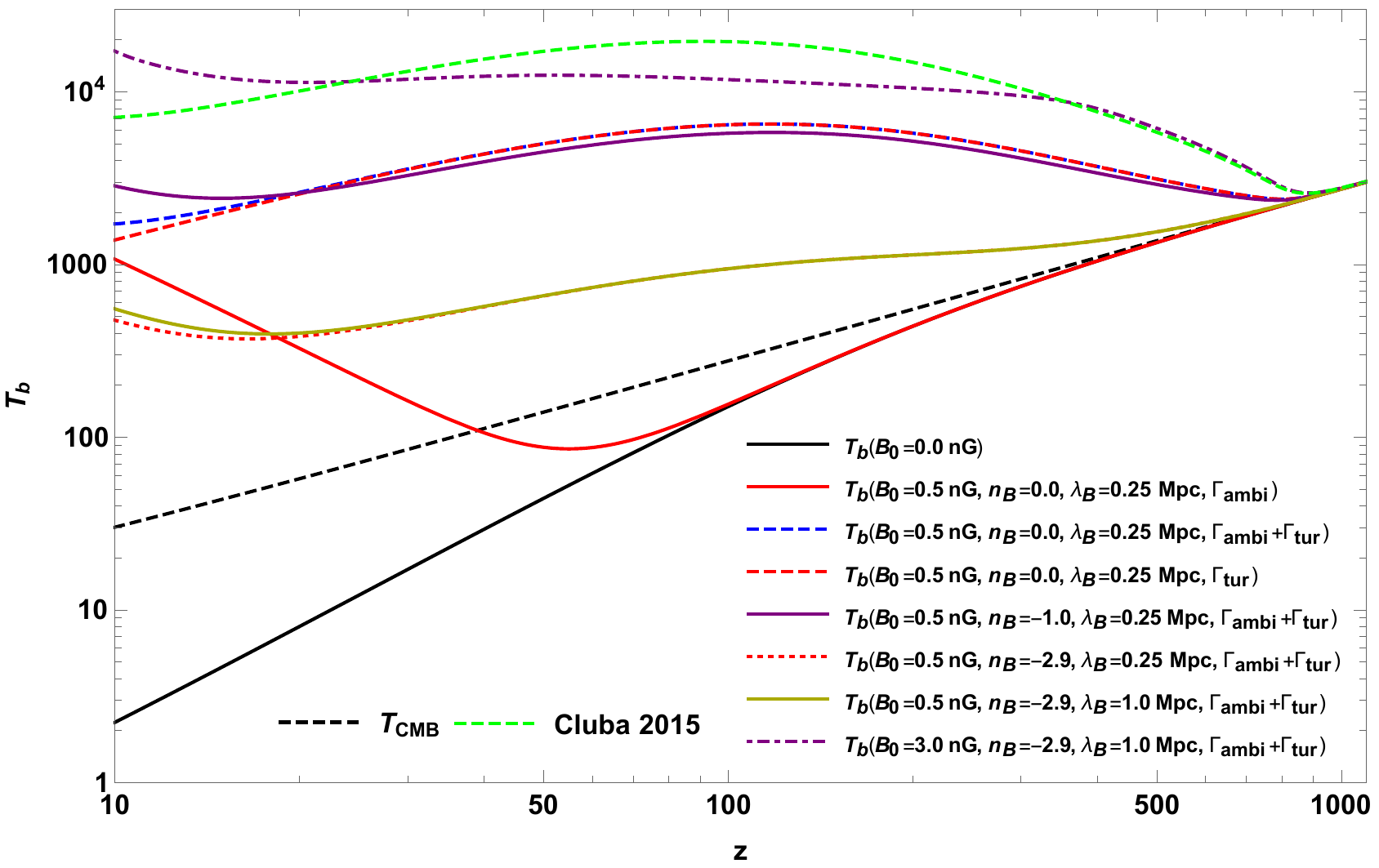}\label{Standard-tem-evolution}}
	\hspace*{1.0 cm}
	\subfloat[]{\includegraphics[width=0.41\linewidth, keepaspectratio]{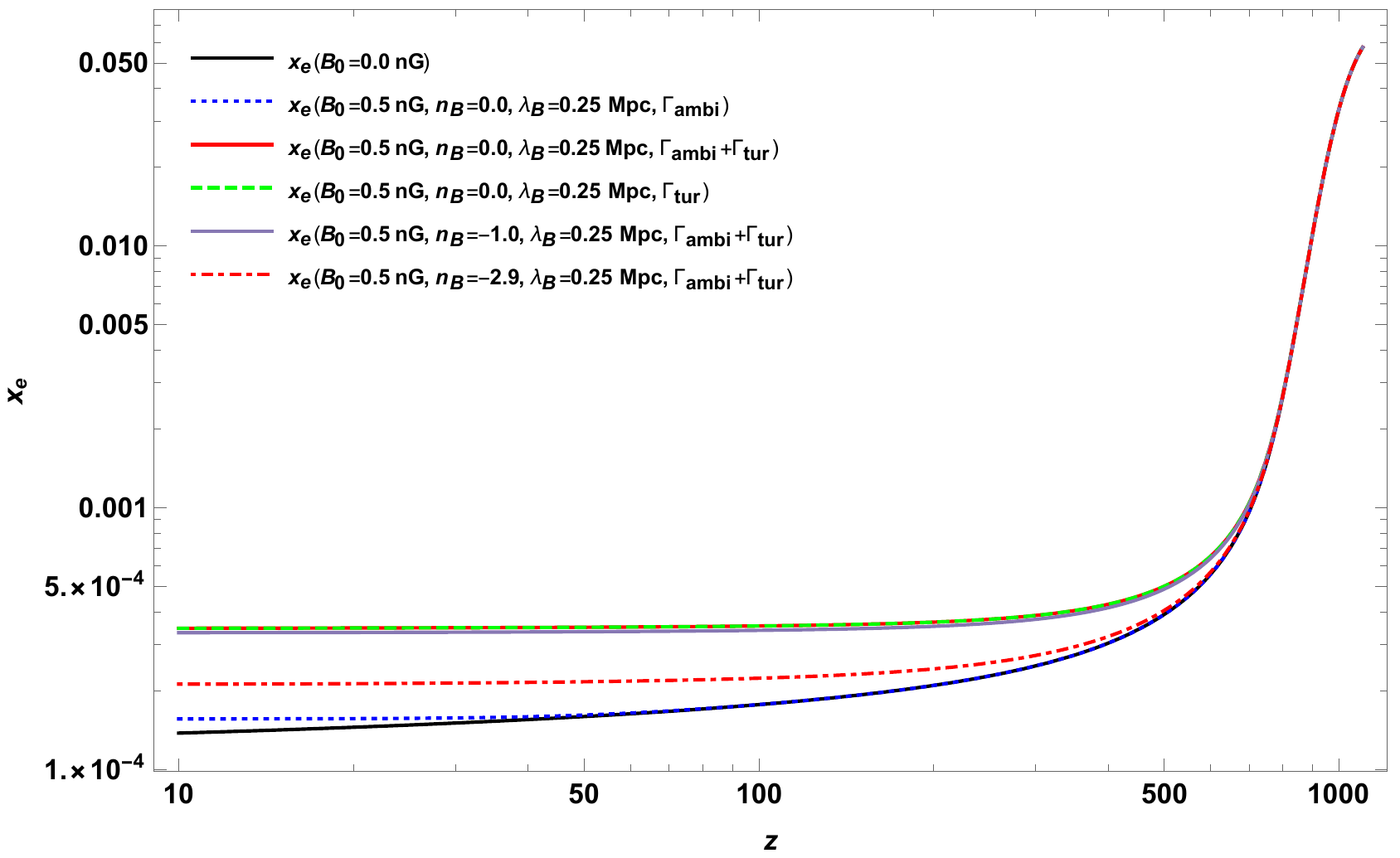}
		\label{standard-xe-evolution}} 
	\caption{\textbf{PMF and no Baryon-DM interaction:} Baryon temperature evolution and ionization fraction evolution with respect to redshift, z is shown in figure (\ref{Standard-tem-evolution}) and (\ref{standard-xe-evolution}), respectively in absence of baryon-DM interactions. Color description is given in the plots.}
	\label{fig:standard-1}
\end{figure*}
\begin{figure*}
	\subfloat[]{\includegraphics[width=0.40\linewidth, keepaspectratio]{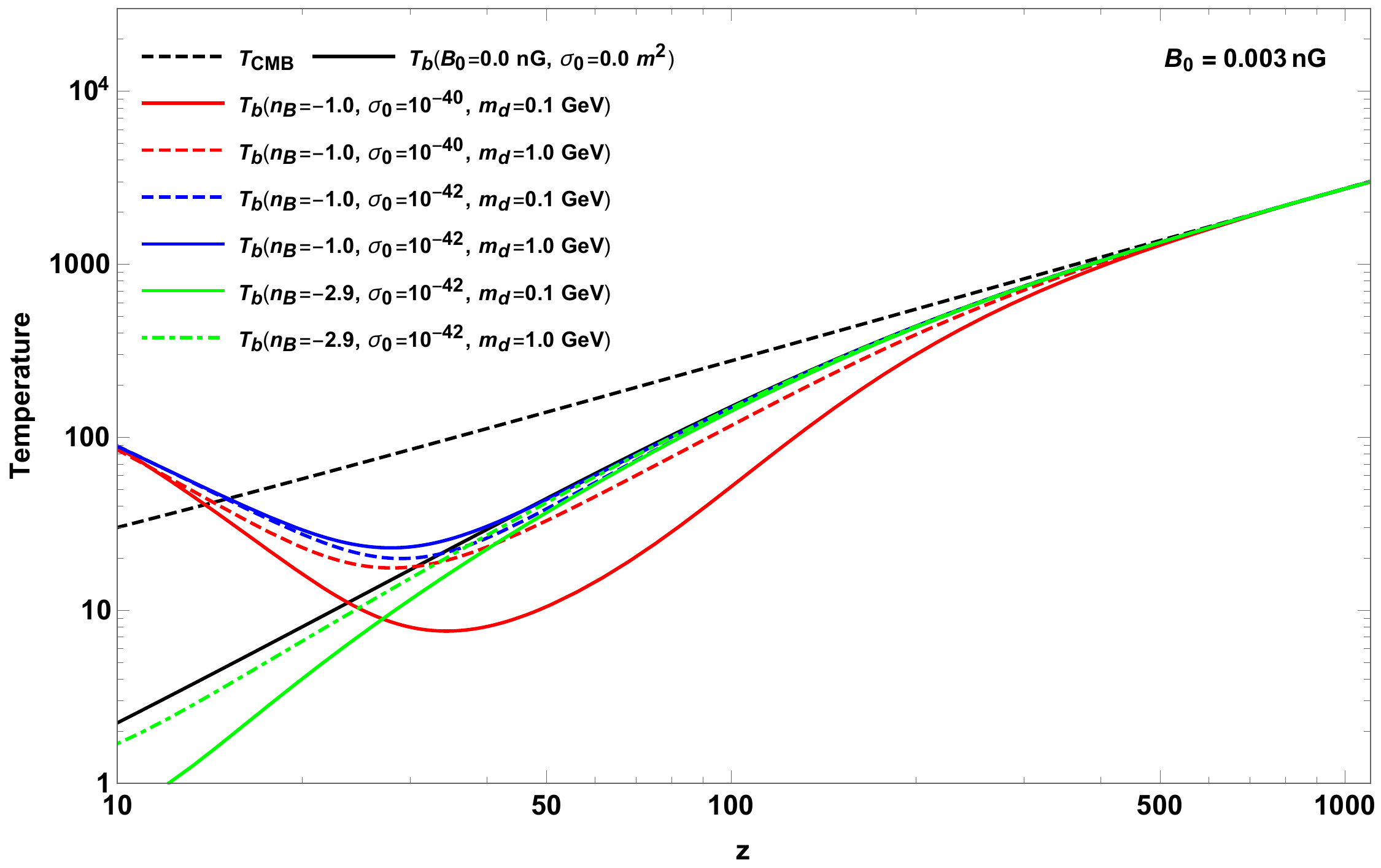}\label{fig:baryon-T-B003}}
	\hspace*{1.0 cm}
	\subfloat[]{\includegraphics[width=0.40\linewidth, keepaspectratio]{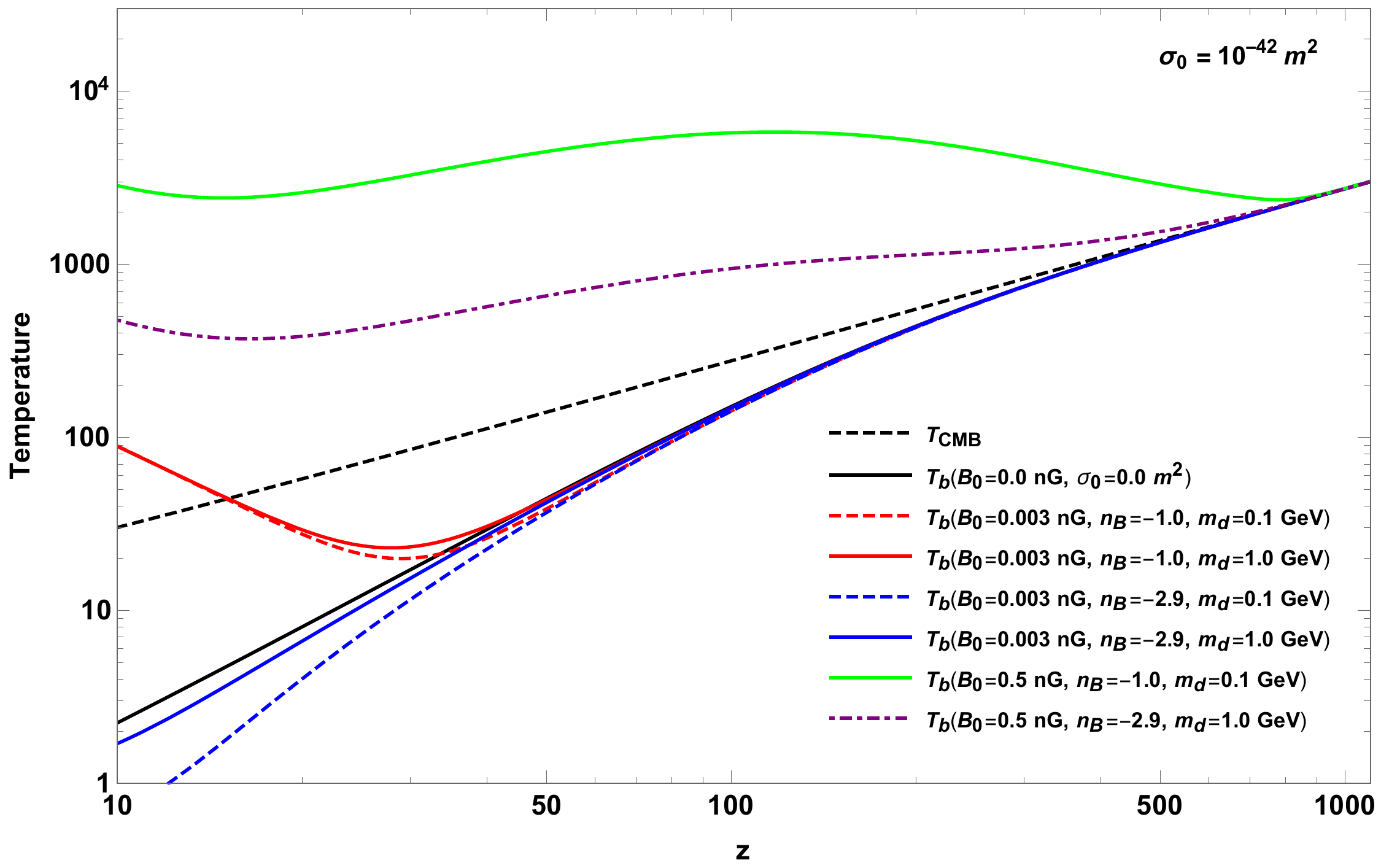}\label{fig:baryon-T-sig-42}} 
	\caption{\textbf{Baryon temperature in presence of Baryon-DM interaction:}
		Evolution of $T_b$ at fixed values of $B_0$ and $\sigma_0$ is shown in figures (\ref{fig:baryon-T-B003}) and (\ref{fig:baryon-T-sig-42}), respectively. Here we have considered $n=-4$.}
	\label{fig:B-DM-Int}
\end{figure*}
\begin{figure*}
	\subfloat[]{\includegraphics[width=0.40\linewidth, keepaspectratio]{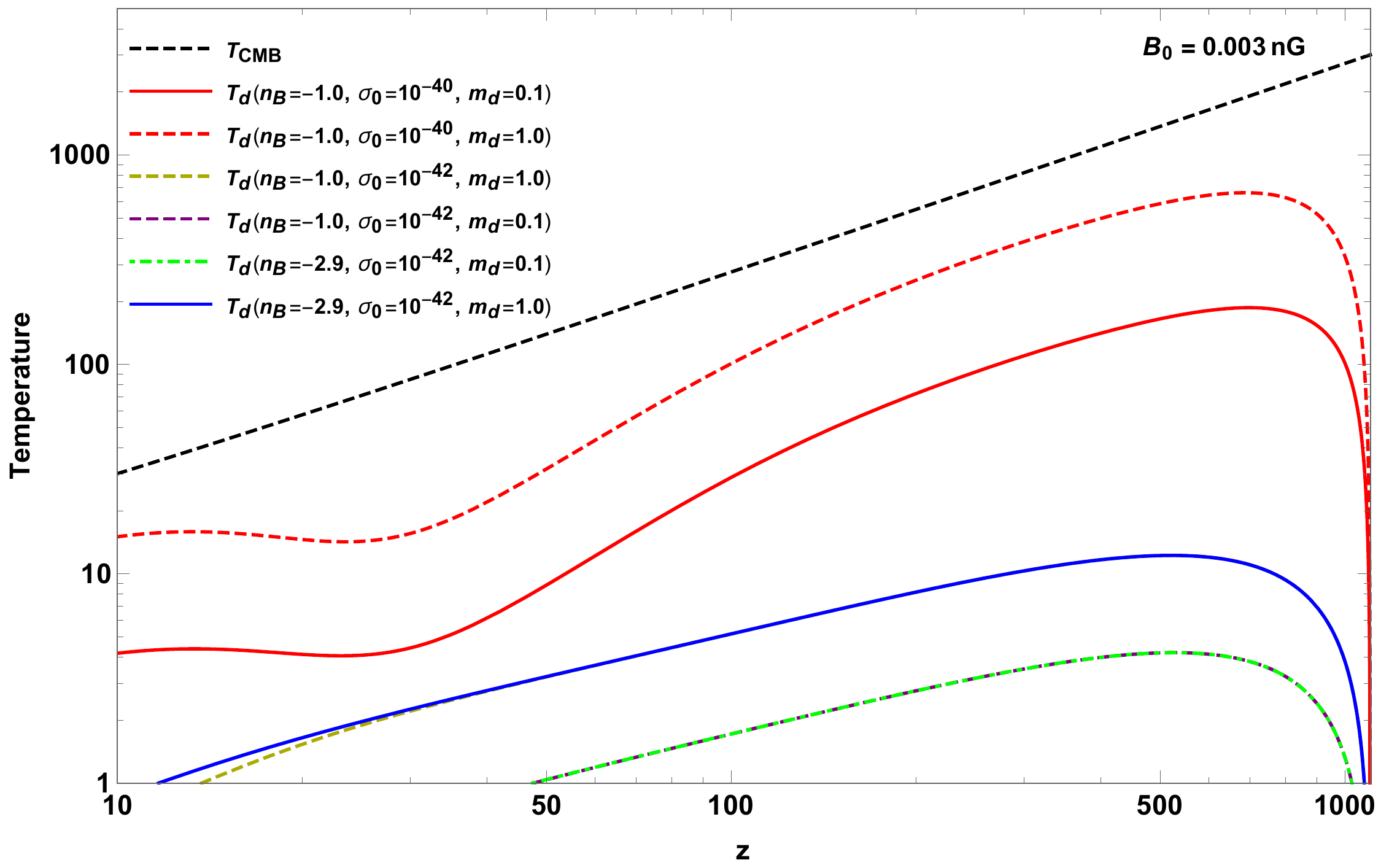}\label{fig:Td-B05}}
	\hspace*{1.0 cm}
	\subfloat[]{\includegraphics[width=0.41\linewidth, keepaspectratio]{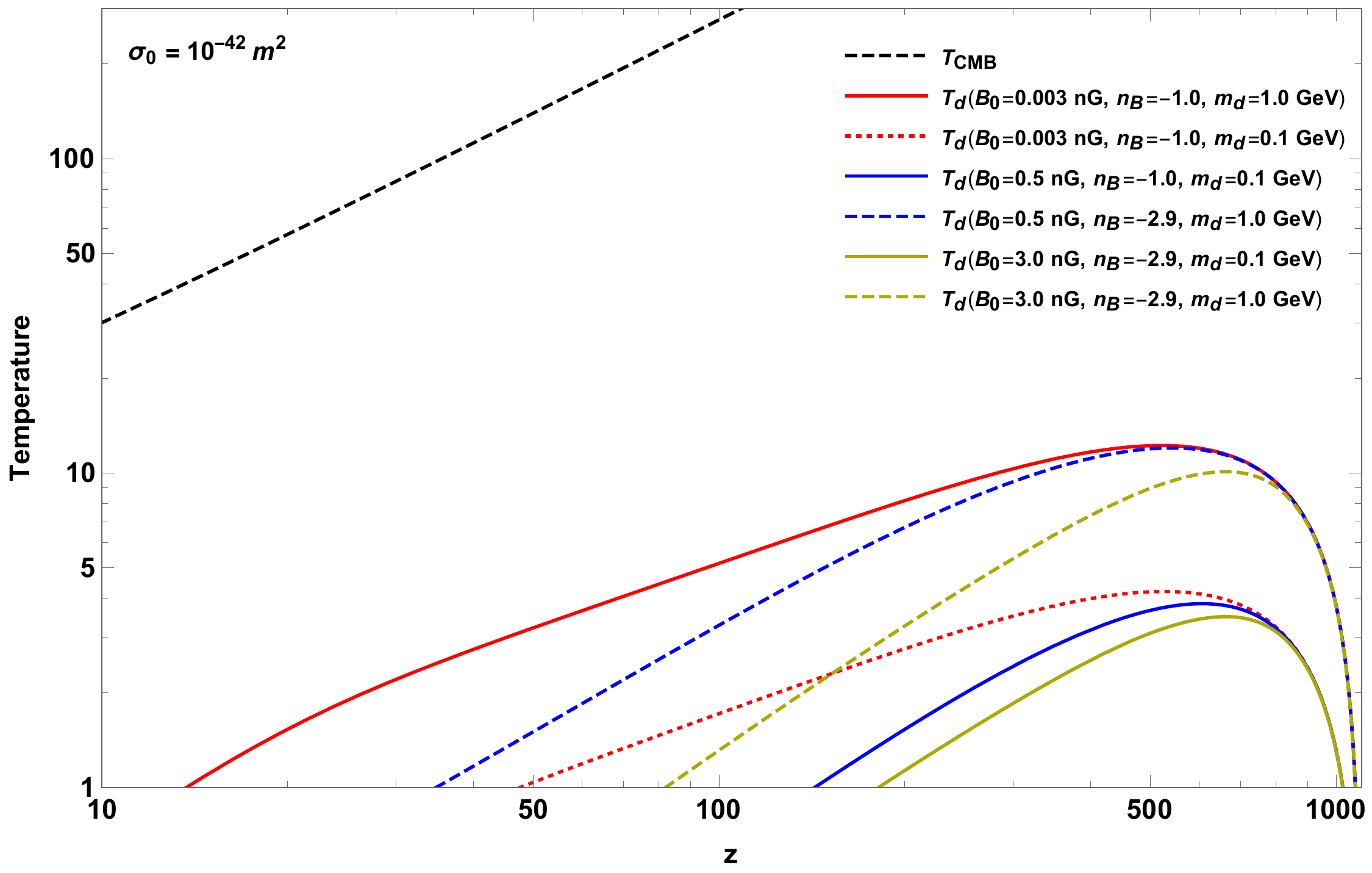}\label{fig:Td-sig042}}
	\caption{\textbf{DM temperature in presence of Baryon-DM interaction:} Evolution of $T_d$ at fixed $B_0$ and at fix $\sigma_0$ is shown in figures (\ref{fig:Td-B05}) and (\ref{fig:Td-sig042}) respectively.  (Here we have considered $n=-4$.)}
	\label{fig:detail}
\end{figure*}
\begin{figure*}
	\subfloat[]{\includegraphics[width=0.40\linewidth, keepaspectratio]{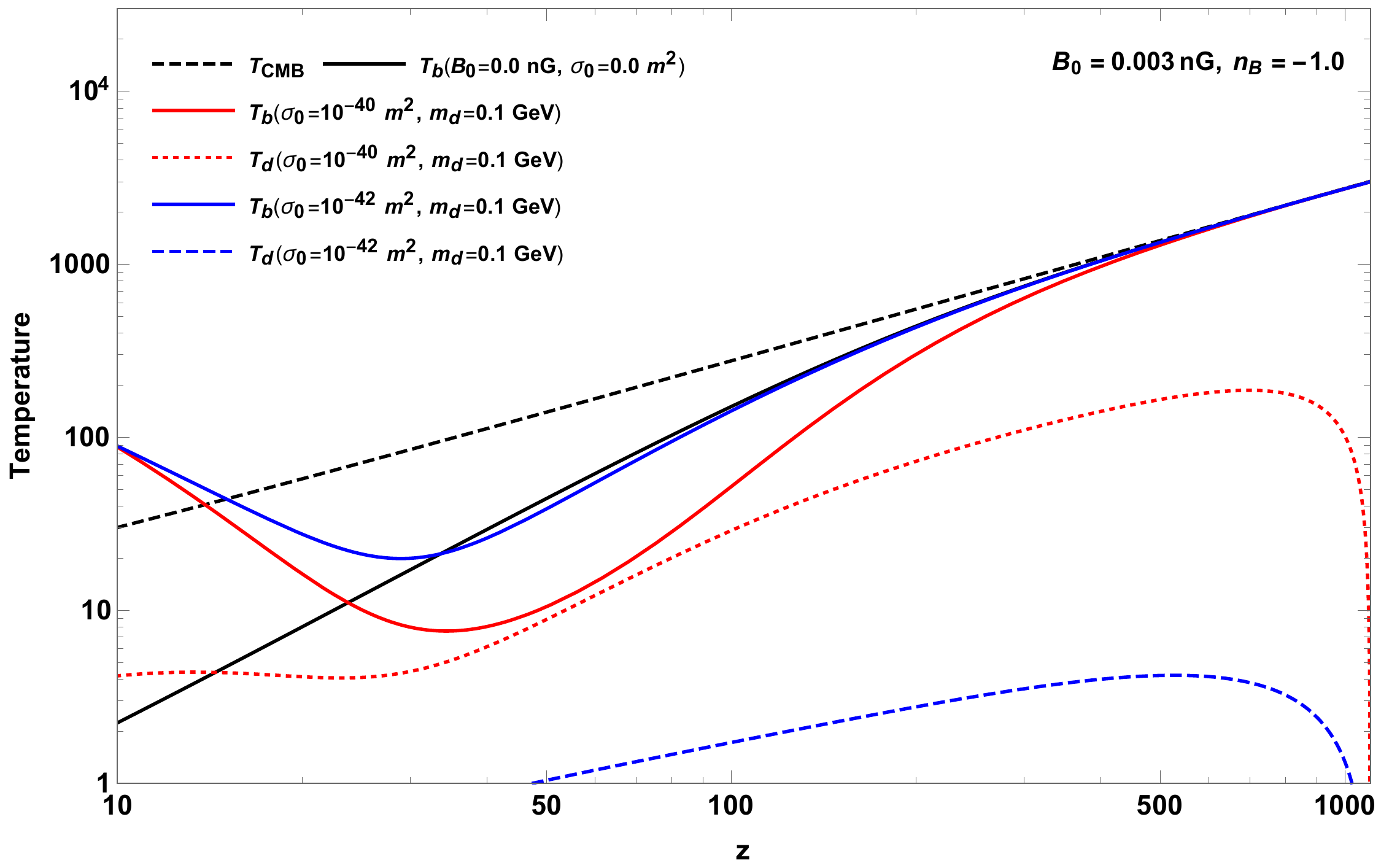}
	\label{fig:Tb-Td-B003-compare}}   
	 \hspace*{1.0 cm}
	\subfloat[]{\includegraphics[width=0.40\linewidth, keepaspectratio]{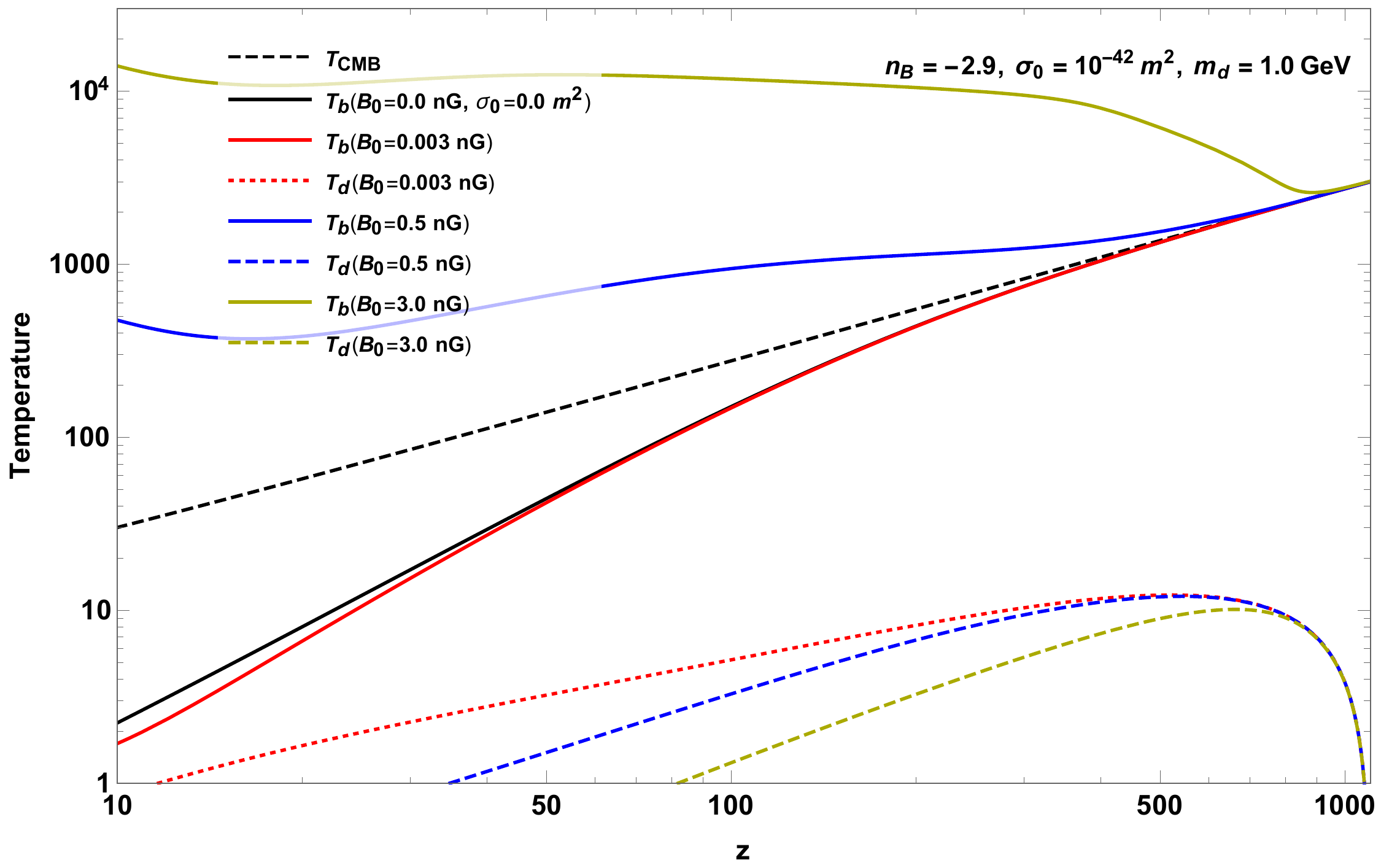}
	\label{fig:Tb-Td-B003-sig042-compare}}
	\caption{\textbf{Baryon-DM temperature comparative study:} Left panel shows the baryon and DM temperature for fix value of magnetic fields strength. In the right panel same as left but  we keep $\sigma_{0}$ and $n_{B}$ to be the same for all the plots but varying magnetic fields strength.}
	\label{fig:compareative-tdtb}
\end{figure*}
\begin{figure}
	\centering
	\includegraphics[width=0.90\linewidth]{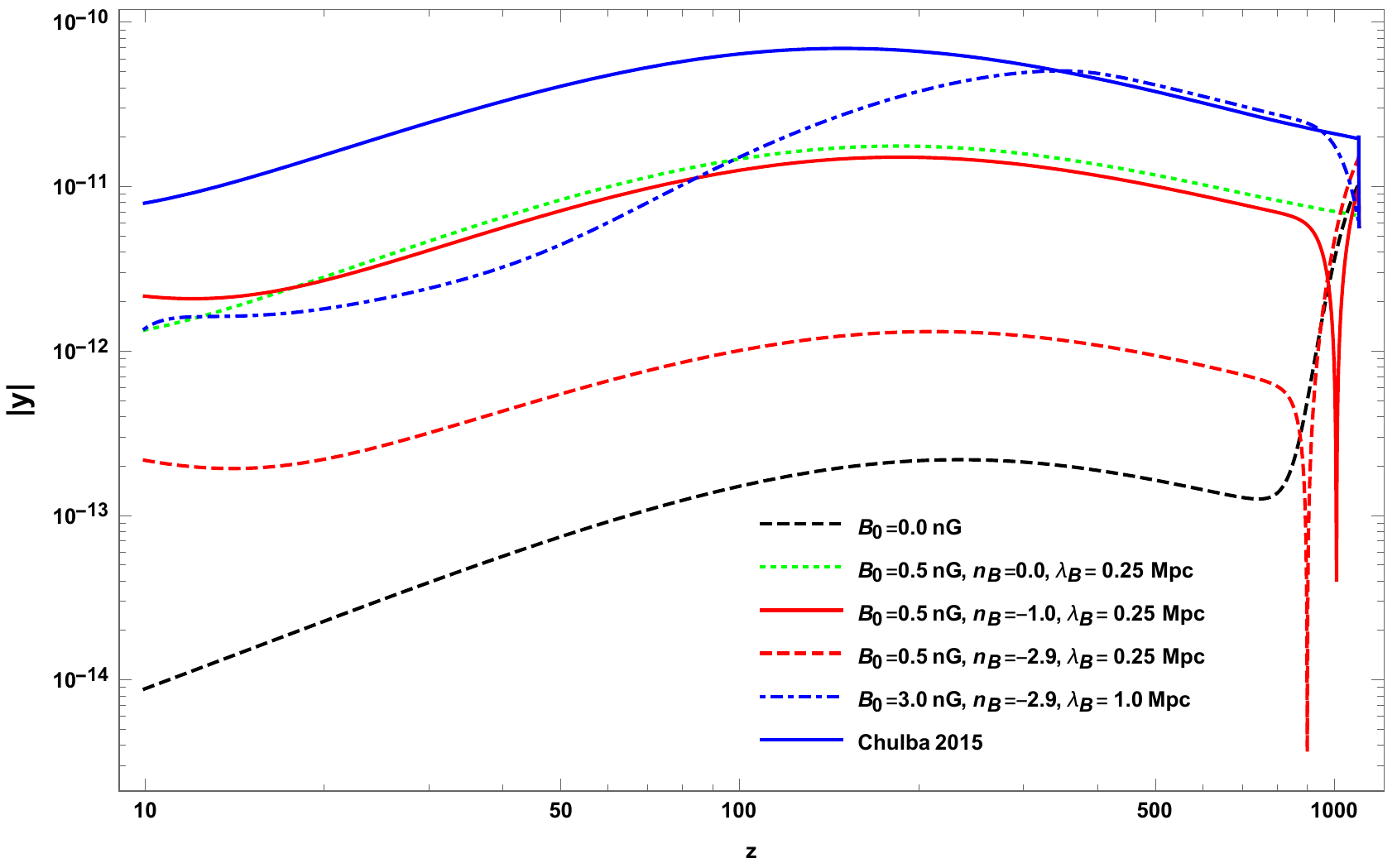}
	\caption{Plot of $|y|$-parameter under no baryon-Dark matter interaction (except gravitational) and with different configurations of magnetic fields.}
	\label{fig:yparam-nodm-1}
\end{figure}
\begin{figure*}
	\subfloat[]{\includegraphics[width=0.40\linewidth, keepaspectratio]{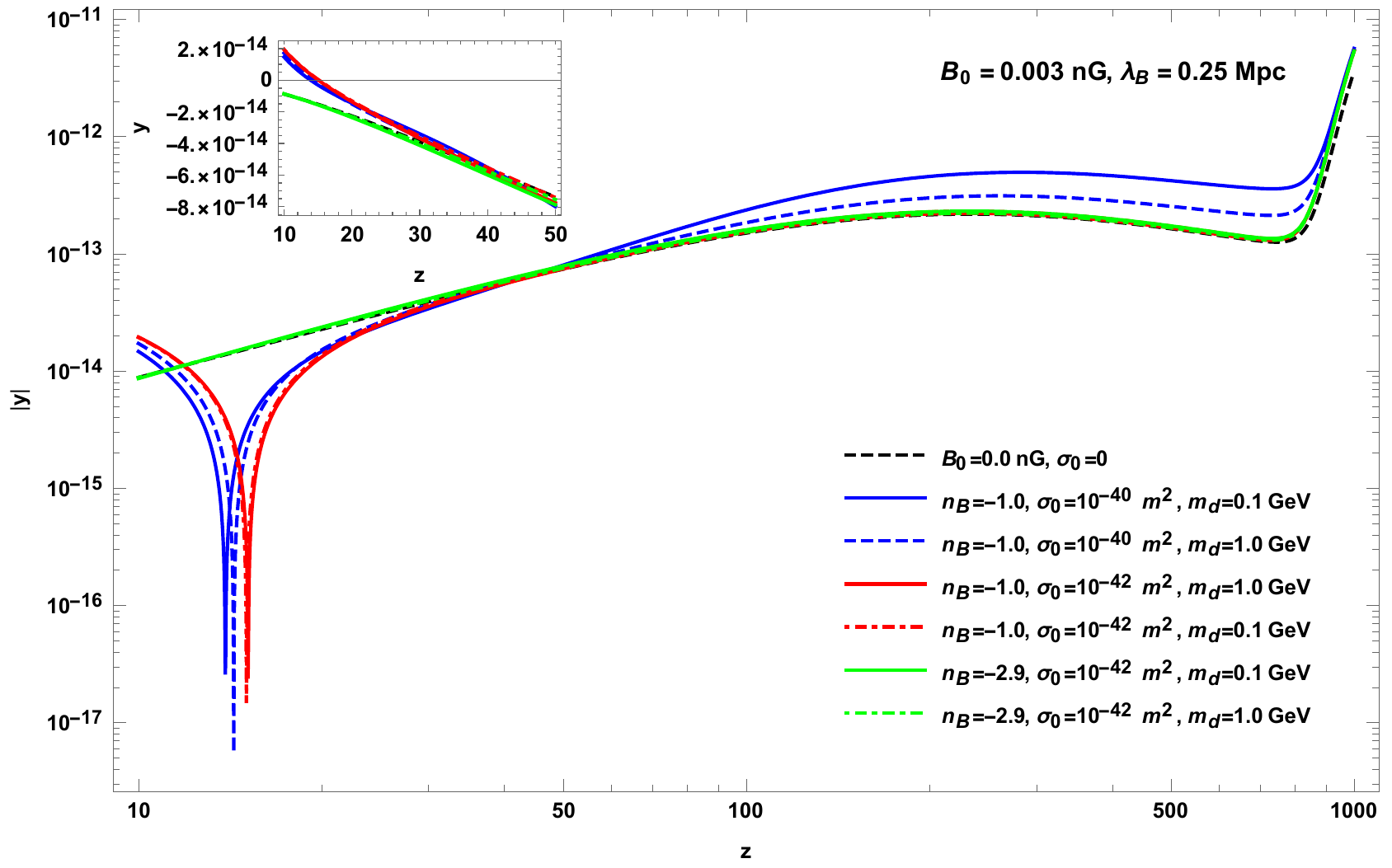}
	\label{fig:y-param-b003}}
	\hspace*{1.0cm}
	\subfloat[]{\includegraphics[width=0.40\linewidth, keepaspectratio]{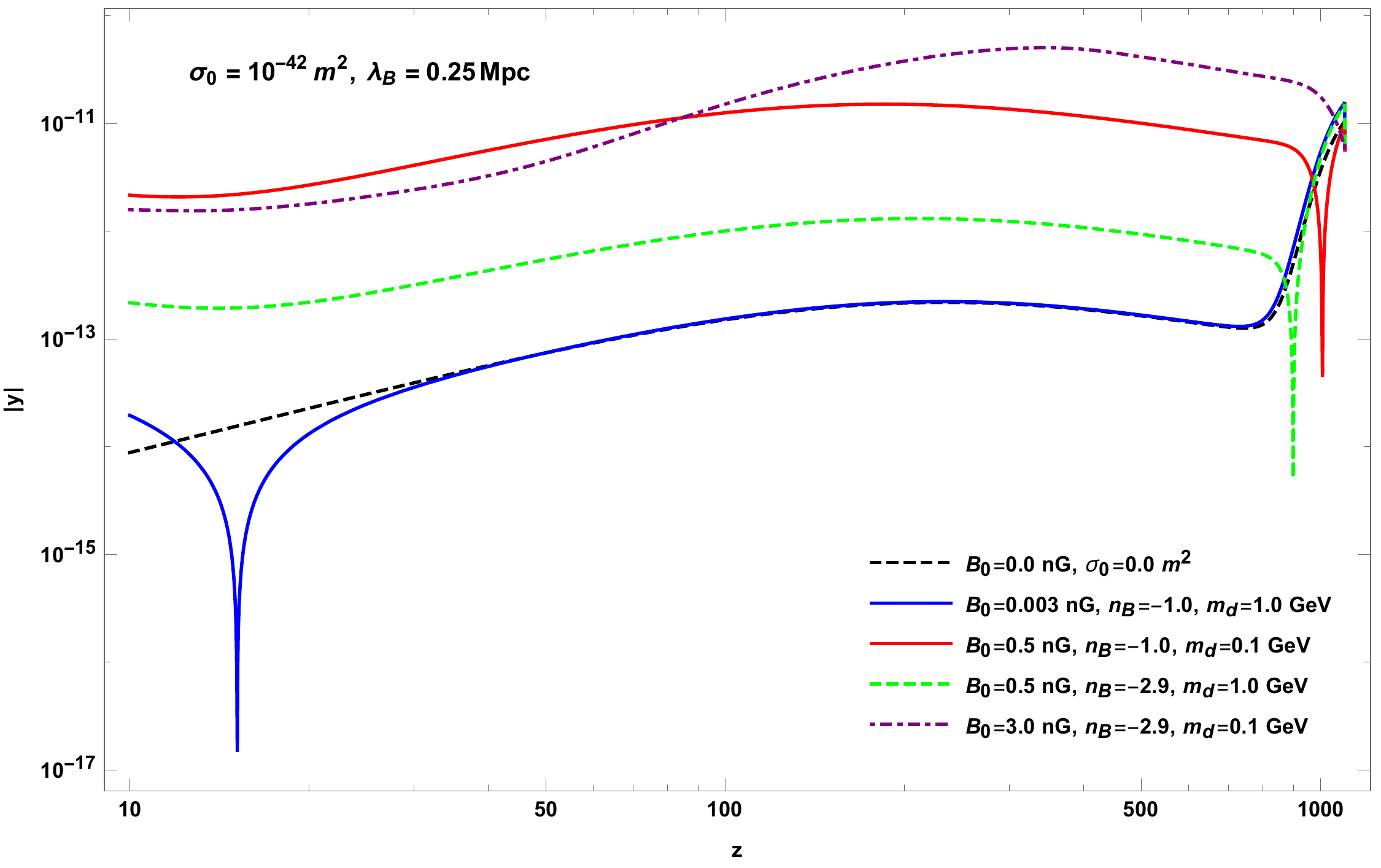}
     \label{fig:y-param-sig42}}
	\caption{\textbf{$|y|$-parameter} for different combination of parameters. In the left panel, the in-set shows the linear plots of the $y-$parameter in redshift range of ($10-50$). The cusp at $z \sim 15$ is due to the reason that at these red shifts, the difference between the baryon temperature and the CMB photons becomes effectively zero for the chosen parameters 
	(see figure (\ref{fig:Tb-Td-B003-compare})).}
	\label{fig:y-param-dm-B}
\end{figure*}
\section{Results \& Discussions}\label{sec-result}
 Impact of the magnetic fields, on the thermal history of the universe after recombination shows that a scale invariant nano-Gauss magnetic field can significantly affect the baryon temperature in IGM (\cite{Chluba:2015lpa, Sethi:2004pe}). In our work, we have considered the thermal effects of magnetic field as well as that of a non-standard interaction between baryons and DM in addition to the effects due to baryon over dense region, cooling from bremsstrahlung process, collisional excitation cooling, the recombination cooling and the collisional ionization cooling. Heat is transferred between the baryons and DM particles due to a finite difference in temperature and drag due to non-zero relative velocity. We would like mention here that we have not included the effect of reionization at redshift $z \leq 10$, as it does not affect our main discussion (\cite{Kunze:2014eka}). Another important aspect of this work is the investigation of the resulting tSZ effect in IGM. We have calculated the implications for the tSZ effect and quantified it under different scenarios. In order to illustrate the interplay of various parameters characterizing the magnetic field and the baryon-dark matter interaction we have considered the cases with mass of $0.1$ GeV and $1.0$ GeV, the interaction cross-sections of $\sigma_0=10^{-40}~\text{m}^2$ and $\sigma_0=10^{-42}~\text{m}^2$, magnetic field strength $B_0$ with values of $0.003$ nG, $0.5$ nG and $3.0$ nG and the coherence length, $\lambda_B$ of $0.25$ Mpc and $1$ Mpc. The values considered for the magnetic spectral index $n_B$ are $0.0$, $-1.0$ and a nearly scale invariant magnetic spectrum with $n_B-2.9$. (for more detail, see Appendix-\ref{sec:mag-spec}). For comparison, we have also taken the case where there is no magnetic field as well as the case where there is no interaction between the baryons and dark matter.
\subsection{Evolution history of interacting dark matter and baryon fluids in presence of magnetic fields}\label{sec:thermal-history}
We have shown the temperature evolution of baryons in the presence of magnetic field but with no BDM interaction in figure (\ref{Standard-tem-evolution}). This is done mainly to use it for comparing with the cases when the BDM interactions are switched on. The CMB temperature is also plotted. The black solid line represents the case where neither the magnetic field nor the BDM interaction is present. When only the ambipolar diffusion for a field strength of $0.5$ nG, spectral index $n_B=0.0$ and a coherence length of $0.25$ Mpc is considered, the effects of the heating the baryons by magnetic field begins to show up clearly at about a redshift of $100$, when it starts getting heated up (solid red line). When both the ambipolar diffusion as well as MHD turbulence is considered, and for the same values for the magnetic field parameters as above, the heating effect shows up much earlier around the redshift of about $900$ itself (blue dashed line). For the case, when $n_B= -1.0$, the effect is very small (red dashed). However, there is more efficient heating at late times for a nearly scale invariant magnetic field ($n_B=-2.9$). By increasing the magnetic field strength and the coherence length scale, the baryon temperature is significantly enhanced (dash dotted purple line) and is more or less consistent with the study of the reference (\cite{Chluba:2015lpa}) with similar configuration (green dashed line). Ionization fraction ($x_e$) is one of the important factor, which tells us about the ionization history of the universe after the recombination epoch. Figure (\ref{standard-xe-evolution}) shows the evolution of $x_e$ with respect to redshift for different combination of the parameters. We have found that the magnetic heating of IGM via ambipolar diffusion and turbulence processes can increase the ionization fraction significantly. One can also notice that the magnetic fields $B_0 = 0.5$ nG with spectral indices, $n_B = 0.0$ has maximum impact on the ionization fraction compared to other parameter values.

We have shown in Figure (\ref{fig:baryon-T-B003}) the results of baryon temperature evolution when there is a BDM interaction in the presence of a weak magnetic field of $0.003$ nG at a magnetic coherence length scale $\lambda_B=0.25$ Mpc. At late times, the magnetic field, although of a strength of $0.003$ nG, heats up the baryons due to the associated dissipation processes of ambipolar diffusion and turbulent decay. Baryon temperature is plotted in figure (\ref{Standard-tem-evolution}). However, in figure (\ref{fig:baryon-T-B003}), since baryons interact with dark matter, the baryons initially cool and hence their temperature decreases. As expected the cooling is faster than the the adiabatic cooling of baryons. At a redshift $z\sim 45$, the heating by magnetic field begins to dominant and the baryon temperature rises. One reason could be a sufficient drop in density of baryons and dark matter, and leads to the effect of BDM interaction becoming insignificant as compared to the heating by magnetic field. At late redshift, they all converge. When we consider a nearly scale invariant magnetic field spectrum $n_B=-2.9$ of the same strength, the effect of magnetic heating does not show up till a redshift of 10. In this case for a BDM interaction cross-section of $\sigma_0=10^{-42}{\text m}^2$, the dashed green curve and the solid green curve are for DM masses of $0.1$ GeV and $1.0$ GeV, respectively. Draining of energy from the baryons to DM is more efficient for a smaller DM mass and hence the smaller the DM mass, more should be the baryon cooling. In fact this is exactly what is reflected in the plots. The fact that a smaller DM mass is more efficient in cooling the baryons is clearly seen  also when $n_B=-1.0$ for the cases of $\sigma_0=10^{-42}~{\text m}^2$ (dashed blue and solid blue curves) as well as for $\sigma_0 = 10^{-40} ~ {\text m}^2$ (dashed red and solid red curves) in figure (\ref{fig:baryon-T-B003}). In addition, for a given BDM interaction strength, we now refer to figure (\ref{fig:baryon-T-sig-42}) and discuss the interplay of magnetic field parameters and the DM mass, for a fixed value of BDM interaction cross-section, for which in our case we have used a value $\sigma_0=10^{-42}$ for illustrative purpose. For a field strength of $B_0=0.003$ nG and DM mass of $m_d=0.1$ GeV the heating is more effective for a spectral index $n_B=-1$ as compared to the case of $n_B=-2.9$ (dashed red and dashed blue curves). A similar trend holds for $m_d=1.0$ GeV (solid red and solid blue curves). For a higher magnetic strength ($B_0=0.5$ nG), increasing the spectral index from $n_B=-2.9$ to $n_B=-1.0$ and decreasing the DM mass from $m_d=1.0$ to $m_d=0.1$ leads to a similar behaviour of a more efficient baryon heating as in the previous case.

We next discuss the resulting temperature evolution of dark matter for various cases in figures (\ref{fig:Td-B05}) and (\ref{fig:Td-sig042}). Since the magnetic field does not directly interact with the dark matter but only through the baryons, there is relatively a small effect on DM temperature by changing the magnetic field parameters while keeping the BDM interaction cross-section and the DM mass same. This can be most clearly seen by the fact that the dashed green curve ($n_B=-2.9$) and the dashed purple curve ($n_B=-1.0$) are almost completely identical. Similarly, for the case of $\sigma_0=10^{-42}~ {\text m}^2$ and $m_d=1.0$ GeV the solid blue curve ($n_B=-2.9$) and the dashed olive green curve ($n_B=-1.0$) almost completely overlap. However, even for the same values of $n_B=-2.9$ and $\sigma_0=10^{-42} ~ {\text m}^2$, the solid blue curve and the dash doted green curve (corresponding to $m_d=1.0$ GeV and $m_d=0.1$ GeV, respectively) are very different. Similarly, for the same value of $n_B=-1.0$ different combinations of $\sigma_0$ and $m_d$ give very different temperature evolution of dark matter, thus emphasizing the fact that it is the BDM interaction parameters which mainly affect the dark matter temperature and that the magnetic field plays a relatively minor role for the chosen parameters.

Furthermore, in figures (\ref{fig:compareative-tdtb}), for comparison, we have described the relative effects on the temperatures of baryons and dark matter for various parameters of the magnetic field and the BDM interaction. The solid blue curve and the dashed blue curve describe the temperature evolution of baryons and dark matter, respectively, for BDM interaction cross-section of $\sigma_0=10^{-42}{\text m}^2$. At early times ($z>600$), dark matter heats up while the baryon cools a bit faster as compared to the adiabatic evolution. At smaller redshifts, both baryons and DM cool along side, but, the latter cools more slowly than the former. Below a redshift of about $50$, the baryons heat up while the dark matter cools to temperatures below $1$ K. This can be explained as follows: due to the fact that while magnetic field heats up the baryons, this dividend is not passed on to the dark matter as the interaction parameters are not enough to offset the effects of lowered density. For $\sigma_0=10^{-40} ~ {\text m}^2$ due to stronger interaction, there is more substantial exchange of energy between baryons and dark matter resulting in a larger spike in dark matter temperature ($z\sim 900$) and larger dip in baryon temperature as compared to the previous case. Again as in the earlier case, the baryon temperature increases below $z\sim 45$ due to magnetic heating and reduced interaction effect and low density. The dark matter will also tend to cool for reasons similar to the previous case. However, as the interaction cross-section is two orders of magnitude more than the previous case, even at these low densities, a certain amount of energy gained by baryons from magnetic decay, is passed on to the dark matter. As a result temperature of the DM remains roughly constant at about $T \sim 7$ K.

Figure (\ref{fig:Tb-Td-B003-sig042-compare}) describes the comparison as in figure (\ref{fig:Tb-Td-B003-compare}) but for different values of magnetic field strength, while keeping spectral index, dark matter mass and the BDM interaction same. We see that as expected, there is relatively small change in the dark matter peak temperature as compared to that of baryons when the magnetic field strength is increased from $0.003$ nG to $3$ nG
(It is useful to keep in mind that the temperature is plotted on a logarithmic scale). While a magnetic field strength of $0.003$ nG is just short of sufficient to offset the additional cooling of baryons due to BDM interaction, field strengths of $0.5$ nG and $3$ nG produce significant rise in baryon temperature.
%
\subsection{SZ-effect and temperature anisotropy} \label{sec:sub-sz}
The temperature and the ionization evolution in presence of BDM and magnetic field, can lead to a distortion in the spectrum of the CMB which is quantified by the $y$-parameter (\cite{Sunyaev:1970ra}). It is given by,
\begin{eqnarray}
y(\hat{n})=\frac{k_B\, \sigma_T}{m_e\, c^2} \int dz \frac{c\, w(\hat{n}, z)}{(1+z) H(z)}, \label{eq:sz-main-equation}
\end{eqnarray}
where $w(\hat{n}, z)=[x_e\, n_b(T_b-T_{\rm CMB})]_{\hat{n}, z}$. The $y$-parameter depends on ionization fraction and the baryon temperature. We have seen that, the baryon temperature and the ionization fraction of the IGM, changes significantly when we include the thermal effect of the magnetic field and the baryon dark matter interactions along with the cooling due to various plasma processes and the density perturbations. It is thus, clear that a small change in baryon temperature and the ionization fraction will create spectral distortion in the CMB via inverse Compton scattering. The variation of the $y$-parameter with respect to the redshift, $z$ is plotted in figures (\ref{fig:yparam-nodm-1}) and (\ref{fig:y-param-dm-B}) in different scenarios of thermal and ionization history in the presence of magnetic field and with baryon-dark matter interaction. Before discussing and analyzing the results shown in these plots, a few points are worth emphasizing. The $y$-parameter can be positive or negative depending on whether the $T_b$ is greater or less than $T_{\rm CMB}$ spectrum. Since the plots are on a log scale, we have plotted $|y|$ instead of $y$. Hence, while interpreting the results, we need to keep in mind that for a given value of $|y|$ the $y$-parameter could be of either sign. The correct sign will be decided on the basis of the value of $(T_b-T_{\rm CMB})$. Further, whenever, the $y$-parameter changes sign, the log$|y|$ plot will show a sharp dip.

Figure (\ref{fig:yparam-nodm-1}) corresponds to different values of magnetic fields but with no BDM interaction. Several points in this figure are worth emphasizing. To begin with if we change the spectral index from $n_B=0.0$ (green dotted) to $n_B=-1.0$ (red solid), the $y$ parameter  undergoes a very small change. However, if we choose a nearly scale invariant spectrum for the magnetic filed ($n_B=-2.9$, red-dashed), the $y$-parameter decreases substantially. To understand the change in the $y$ parameter, this figure needs to be analyzed in conjunction with figure (\ref{Standard-tem-evolution}). In this figure we see that when all the effects of the magnetic field are taken into account, the baryon temperature is always more than the CMB temperature. This tends to give a positive $y$-parameter. In figure (\ref{Standard-tem-evolution}), the only case where the baryon temperature is below CMB temperature and later moves above the CMB temperature, is the case when we neglect the contribution due to turbulence. If we, however, neglect the magnetic field ($B_0= 0.0$ nG), we see from figure (\ref{Standard-tem-evolution}) that the baryon temperature is below CMB temperature. Hence, the $y$-parameter is negative. In Figure (\ref{fig:yparam-nodm-1}), it should be noted that we have plotted the $|y|$ (black dashed line). When we keep this in mind we see that with the increase of $B_0$, the general trend is that the $y$-parameter increases. When $B_0=3.0$ nG, the behavior of the evolution curve of $y-$parameter is slightly different for a nearly scale invariant spectrum and a coherence length of $1$ Mpc. To compare our results obtained, we have also included a plot for the $y-$ parameter for a similar configuration of ref. (\cite{Chluba:2015lpa} ($B_0=3.0$ nG, $n_B=-2.9$ and $\lambda_B=1.0$ Mpc). Since they do not consider the effects of cooling and density perturbation sourced by magnetic fields, the $y$-parameter in their case is higher for redshifts less than about redshift $z\sim$ 400 although the other parameter values are the same.

The effect of switching on the BDM interaction along with magnetic fields and density perturbations and cooling effects, is shown in Figure (\ref{fig:y-param-dm-B}). The behavior of the evolution of the $y$ parameter for a fixed magnetic field of $0.003$ nG and a coherence length of $0.25$ Mpc and with different values of interaction cross section, mass of the DM particles and the magnetic field spectral index are shown in figure (\ref{fig:y-param-b003}). The case of zero magnetic field (black dashed line) is also shown for comparison. We discuss the results shown in figure (\ref{fig:y-param-b003}) again in conjunction with figure (\ref{fig:baryon-T-B003}). Those cases in figure (\ref{fig:baryon-T-B003}), where the $T_b$ is always less than $T_{\rm CMB}$, will cause the $y$-parameter to be negative. As shown in figure (\ref{fig:y-param-sig42}), the behavior of log$|y|$ is smooth as shown. However, for other cases log$|y|$ shows sharp a dips at red shifts between $20$ to $40$ depending on the model parameters. In order to understand this, let us focus on the inset where $y$ is plotted with respect to $z$ on a linear scale (without taking modulus) in the redshift range $10$ to $50$. We see that the $y-$parameter is initially negative and crosses over to positive values. This can again be understood in terms of the baryon temperature evolution in figure (\ref{fig:baryon-T-B003}). For the values $B_0 \geq 0.5$ nG the baryon temperature is initially below CMB temperature and crosses over around a redshift of $20$ to $40$. It is precisely at this redshift, where the $y$-parameter in these models cross-over to positive values. In the main figure the sharp dip in the value of of log$|y|$ is when $y$ is zero at the cross over redshift. On the right of it, $y$ is negative and on the left it is positive. Another observation worth noting is that with all other parameters being the same, the curves are relatively insensitive to the value of the mass when we vary it from $0.1$ GeV to $1.0$ GeV.
\section{Conclusion}\label{sec-conl}
The nature of the evolution of the thermal and ionization history of the IGM depends on the relative strength and nature of the magnetic field, (and its heating arising out of ambipolar diffusion and turbulent decay) on one hand, and the cooling of baryons due the baryon-DM interaction on the other. This in turn leave imprints through tSZ effect on the CMB distortion at small-scales. The latter is quantified by a non-zero y-parameter. In this analysis, we have considered BDM interaction for $n = -4$. While there is physical motivation for the cases of $n=-2$ and $n=-1$, these do not lead to any significant impact on our analysis and hence have not been considered in this work.
\begin{itemize}
    \item The baryon temperature tends to increase with magnetic field strength. Magnetic turbulence has prominent role in governing the baryon temperature along with other processes. The results strongly depend on the magnetic spectral indices if it is nearly scale invariant. Otherwise the dependence was found to be  relatively weak.
    \item The ionization fraction increases with the strength of magnetic fields. The energy transfer to baryons due to ambipolar and turbulent decay of magnetic field, contributes to the ionization.
    \item The interaction cross-section between the Baryons and Dark matter has considerable impact of the cooling of the baryon and heating of Dark Matter. Further, the mass of the dark matter particles have a relatively weak influence on the  thermal history of the baryons.
    \item When the magnetic field is weak, its heating influence becomes significant only at later time (smaller redshifts). In these scenarios the baryon cooling starts sooner in the post recombination era. With expansion their density and hence the effect of their interaction drops and even the weak magnetic field catches up and heats the baryons subsequently. For relatively stronger magnetic fields, on the other hand, the heating starts at fairly large redshifts and the baryon temperature does not go below the CMB temperature.
    \item The above effects show a similar behaviour in the evolution of the $y$-parameter with redshift. For weak magnetic fields, the crossover of the baryon temperature $T_{\rm CMB}$ being less than CMB temperature and being more than it later, results in the $y-$parameter becoming $+$ve at early times and it becomes positive later times.
\end{itemize}
In summary, we have investigated the effect of magnetic heating of baryons in
IGM in presence/absence of BDM interaction. Our investigation highlights the difference in thermal history of baryons in these two cases. Further, we have also studied the evolution of resultant spectral distortion, qualified by $y$-parameter.
\appendix
\section{Statistical properties of the magnetic fields}\label{sec:mag-spec}
Let us consider a homogeneous and isotropic Gaussian magnetic field. The statistical properties of these fields can be obtained through the two point correlation functions $\langle B_i({\bf x})B_j({\bf x}+{\bf r})\rangle =B_{ij}({\bf r})$ (where $\langle ...\rangle$ denotes the average over the statistical ensemble) is defined as (\cite{Monin:1971sa, Durrer:2003ja}) 
\begin{equation}
B_{ij}({\bf r})= P_T (r)\delta_{ij}+[P_L(r)-P_T(r)]\,  \hat{\bf r}_i \hat{\bf r}_j + P_A (r)\, \epsilon_{ijk}r_k.
\end{equation}
Here $P_T$, $P_L$ and $P_A$ are transverse, longitudinal and helical (antisymmetric) components of the magnetic field correlation function respectively. $\hat{\bf r}= r_i/|{\bf r}|$. For antisymmetric helical fields, rotational symmetry is preserved and hence $B_{ij}(-{\bf r})=B_{ij}({\bf r})$. The three components can be obtained by following relations
\begin{eqnarray}
P_T(r) & = & \frac{1}{2}B_{ij}({\bf r})\left(\delta_{ij}-\hat{\bf r}_i \hat{\bf r}_j\right)\nonumber \\
P_L(r) & = & B_{ij}({\bf r}) \,\hat{\bf r}_i \hat{\bf r}_j \nonumber \\
P_A(r)& = & \frac{1}{2r}\, B_{ij}({\bf r})\epsilon_{ijk} \hat{\bf r}_k
\end{eqnarray}
In terms of spectral i.e. Fourier decomposition of the stochastic magnetic field amplitudes ${\bf B}({\bf k})$, the two point correlation defined above is written as:
\begin{eqnarray} \label{eq:twopoint-corr}
\langle {\bf B}_i^*({\bf k})\, {\bf B}_j({\bf k}')\rangle= (2\pi)^3 \, \delta^3({\bf k}-{\bf k}')\, \mathcal{F}_{ij}^B({\bf k})\, , 
\end{eqnarray}
where $\langle ..... \rangle$ denotes ensemble average in Fourier space. 
We define the Fourier transform of the magnetic field, $B({\bf x})$ as 
\begin{eqnarray}
{\bf B}({\bf x})& = & \frac{1}{(2\pi)^3}\int d^3\, {\bf k}\, \exp(-i\, {\bf k}\cdot {\bf x})\,   {\bf B}({\bf k}) \nonumber \\
{\bf B}({\bf k})& = & \int d^3\, {\bf x}\, \exp(i\, {\bf k}\cdot {\bf x})\,   {\bf B}({\bf x}) \nonumber
\end{eqnarray}
Here $\mathcal{F}_{ij}^B({\bf k})$ is defined as
\begin{eqnarray}
B_{ij}({\bf r}) & = & \frac{1}{(2\pi)^3}\int d^3{\bf k}\, \exp(-i{\bf k}\cdot {\bf r})\, \mathcal{F}_{ij}^B({\bf k}),  \nonumber \\
\mathcal{F}_{ij}^B({\bf k}) & = & \int d^3 {\bf r}\,  \exp(i{\bf k}\cdot {\bf r})\, B_{ij}({\bf r})  .
\end{eqnarray}
For a helical magnetic field, function $\mathcal{F}_{ij}^B({\bf k})$ satisfies 
\begin{eqnarray}
\mathcal{F}_{ij}^B({\bf k})=\mathcal{F}_{ij}^B(-{\bf k})=[\mathcal{F}_{ij}^B]^*(-{\bf k})=[\mathcal{F}_{ij}^B]^*({\bf k})
\end{eqnarray}
Similar to the real space function $B_{ij}({\bf r})$, we can write $\mathcal{F}_{ij}^B({\bf k})$ in following form
\begin{eqnarray}
\frac{\mathcal{F}_{ij}^B({\bf k})}{(2\pi)^3}=P_{ij}(\hat{{\bf k}}) \frac{\mathcal{P}_{B}(k)}{4\pi k^2} + i \epsilon_{ijl} k_l \frac{\mathcal{H}_{B}(k)}{8\pi k^3}\, ,
\end{eqnarray}
where $\mathcal{P}_{B}$ and $\mathcal{H}_{B}$ are the symmetric and antisymmetric part of the two point correlation of the magnetic fields in the Fourier space. 
The PMF power spectrum is defined as the Fourier
transform of the two point correlation defined in equation (\ref{eq:twopoint-corr}). The projection operator $P_{ij}=\delta_{ij}-\hat{\bf k}_i\, \hat{{\bf k}}_j$ project $\mathcal{F}_{ij}$ onto the transverse plane and $\epsilon_{ijl}$ is the 3D Levi-Civita tensor, where $P_{ij}\, \hat{k}^i=0$. The symmetric and antisymmetric part of the above two point correlations are given by:
\begin{eqnarray}
\langle {\bf B}_i^*({\bf k})\, {\bf B}^*_i({\bf k}')\rangle & = & 2 (2\pi)^3 \delta({\bf k}-{\bf k}')  \mathcal{P}_{B}(k) \nonumber \\
-i  \langle \epsilon_{ijk} \hat{{\bf k}}^k\,{\bf B}_i^*({\bf k})\, {\bf B}^*_j({\bf k}')\rangle & = & 2 (2\pi)^3 \delta({\bf k}-{\bf k}') \mathcal{H}_{B}(k).
\end{eqnarray}
For the case of power law magnetic spectra, i.e. $\mathcal{P}_{B}=A\, k^{n_B}$ and $\mathcal{H}_{B}=A_H\, k^{n_H}$, for $k< k_D$ and  $\mathcal{P}_{B}=0$, $\mathcal{H}_{B}=0$ for $k>k_D$, the amplitude $A_B$ and $A_H$ is given by (\cite{Trivedi:2012ssp, Trivedi:2013wqa, Hortua:2015jkz, Ade:2015cva})
\begin{eqnarray}
A= \frac{(2\pi)^2\, 2^{(n_B+1)/2}\, \langle B^2\rangle_\lambda}{\Gamma \left(\frac{n_B+3}{2}\,  \right)\, k_\lambda^{3+n_B}}, \\
A_H= \frac{(2\pi)^2\, 2^{(n_H+1)/2}\, \langle \mathcal{H}^2\rangle_\lambda}{\Gamma \left(\frac{n_H+4}{2}\,  \right)\, k_\lambda^{3+n_H}}, \, , 
\end{eqnarray}
where $B_\lambda$ and $\mathcal{H}_\lambda$ are the comoving magnetic field strength and magnetic helicity smoothed over a Gaussian sphere of comoving radius $k_D$. The smoothing function that we have used is $~\exp(-2k^2/k_D^2)$. The smoothed magnetic fields over a comoving scale $k< k_D$ evolves linearly in post-recombination epoch. To avoid infrared divergence, $n_B$ and $n_H$ should satisfy $n_B>-3$ and $n_H>-4$. At this stage dissipates mainly by ambipolar diffusion.  However, in the most general case, a infrared cut off at $k_{\rm inf}$ should be considered for the power spectrum of the magnetic and the helical energy in the range of $k_{\rm inf}\leq k\leq k_D$ (where $k_{\rm inf}$ infrared cut off scale and $k_D$ is the ultraviolet cut off scale and correspondence to damping scale where the field is suppressed on small scales (\cite{Dolag_2010})). At length scales $k_D\leq k\leq k_{\rm max}$, non-linear effects can leads to decaying MHD turbulence and in result dissipation of the magnetic energy. In the post-recombination epoch, density perturbations seeded by the magnetic fields can grow at scale $k\leq k_D$, can leads to early structure formation \cite{Wasserman:1978iw, Kim:1994zh, Sethi:2004pe}.
\section*{Acknowledgments}
	AKP, SM and TRS acknowledge the facilities at I.C.A.R.D, University of Delhi. AKP is supported by the Dr. D.S. Kothari Post-Doctoral Fellowship provided by Govt. of India, under the Grant No. DSKPDF Ref. No. F.$4-2/2006$ (BSR)/PH/$18-19/0070$. The research of SM is supported by UGC, Govt. of India under the UGC-JRF scheme (Sr.No. $2061651305$ Ref.No: $19/06/2016$(I) EU-V). TRS acknowledge the project grant from SERB, Govt. of India (EMR$/2016/002286$). AKP would also like to thank IUCAA, Pune and SM also acknowledge NCRA, Pune for providing hospitality during the visit, where some part of this work was done.
\bibliographystyle{mnras}
\bibliography{mnrasref} 
\label{lastpage}
\end{document}